\documentclass[aps,prd,twocolumn,preprintnumbers,amsmath,amssymb,superscriptaddress,longbibliography,nofootinbib]{revtex4-1}
\usepackage{graphicx}
\usepackage{dcolumn}
\usepackage{bm}
\usepackage{color}
\usepackage{ulem}
\usepackage[usenames,dvipsnames]{xcolor}
\usepackage{hyperref}

\hypersetup{
    bookmarks=false,        
    colorlinks=true,        
    linkcolor=blue,        
    citecolor=blue,        
    filecolor=blue,      
    urlcolor=blue
}

\newcommand{\ket}[1]{|#1\rangle}

\newcommand{\GFzub}[2]{\langle\!\langle #1|#2\rangle\!\rangle}

\newcommand{\GF}[2]{\langle \{ #1,#2 \}\rangle}

\newcommand{\be}{\begin{equation}}
\newcommand{\ee}{\end{equation}}
\newcommand{\bea}{\begin{eqnarray}}
\newcommand{\eea}{\end{eqnarray}}
\newcommand{\eq}[1]{Eq.~(\ref{#1})}
\newcommand{\fig}[1]{Fig.~\ref{#1}}
\newcommand{\figs}[1]{Figs.~\ref{#1}}
                             
\newcommand{\e}{\varepsilon}
\newcommand{\w}{\omega}
\newcommand{\s}{\sigma}
\newcommand{\G}{\Gamma}
\newcommand{\up}{\uparrow}
\newcommand{\down}{\downarrow}

\renewcommand{\S}{\mathcal{S}}

\newcommand{\dk}{d^\dagger}
\newcommand{\fk}{f^\dagger}

\newcommand{\upx}{\!\uparrow\!}
\newcommand{\downx}{\!\downarrow\!}
\newcommand{\nn}{\nonumber}
\newcommand{\Jeff}{J_{\rm eff}}

\begin{document}

\title{Majorana-Kondo interplay in T-shaped double quantum dots}

\author{I. Weymann}
\email{weymann@amu.edu.pl}
\affiliation{Faculty of Physics, Adam Mickiewicz University, 
		ul.~Uniwersytetu Pozna\'{n}skiego 2, 61-614 Pozna\'{n}, Poland}
		
\author{K. P. W\'ojcik}
\email{kpwojcik@ifmpan.poznan.pl}
\affiliation{Institute of Molecular Physics, Polish Academy of Sciences, ul.~Smoluchowskiego 17, 60-179 Pozna{\'n}, Poland}
\affiliation{Physikalisches Institut, Universit\"at Bonn, Nussallee 12, D-53115 Bonn, Germany}

\author{P. Majek}
\affiliation{Faculty of Physics, Adam Mickiewicz University, 
		ul.~Uniwersytetu Pozna\'{n}skiego 2, 61-614 Pozna\'{n}, Poland}

\date{\today}

\begin{abstract}
The transport behavior of a double quantum dot side-attached to a topological
superconducting wire hosting Majorana zero-energy modes is studied theoretically 
in the strong correlation regime. 
It is shown that Majorana modes can leak to the whole nanostructure,
giving rise to a subtle interplay between the two-stage Kondo screening and the 
half-fermionic nature of Majorana quasiparticles. 
In particular, the coupling to the topological wire is found to reduce the effective
exchange interaction between the two quantum dots in the absence of normal leads.
Interestingly, it also results in an enhancement of the second-stage Kondo 
temperature when the normal leads are attached. Moreover, it is shown that the second 
stage of the Kondo effect can become significantly modified in one of the spin channels 
due to the interference with the Majorana zero-energy mode, yielding the low-temperature
conductance equal to $G=G_0/4$, where $G_0 = 2e^2/h$,
instead of $G=0$ in the absence of the topological superconducting wire.
We also identify a nontrivial spin-charge ${\rm SU}(2)$ symmetry present in the system 
at a particular point in parameter space, despite lack of the spin nor charge conservation.
Finally, we discuss the consequences of a finite overlap between two Majorana modes, 
as relevant for short Majorana wires.
\end{abstract}

\maketitle

\section{Introduction}

Topological states of matter are
in the center of current research in condensed matter physics
\cite{Hasan2010Nov,Qi2011Oct,Wang2017Oct}.
This is due to the fact that such states
are robust against decoherence and
are thus very promising for applications
in quantum information and computation \cite{Nayak2008Sep}.
In this regard, topologically-protected states that form
at the ends of one-dimensional topological superconductor,
referred to as Majorana zero-energy modes \cite{Majorana1937Apr},
provide an exciting example \cite{Kitaev2003Jan,Alicea2012Jun}.
The signatures of such Majorana quasiparticles
have been recently reported in a number of experiments
\cite{Mourik2012May,Deng2012Nov,Das2012Nov,Albrecht2016Mar,
Deng2016Dec,Deng2018Aug,Zhang2018Mar,Lutchyn2018May,Gul2018Jan}.

It has been demonstrated that
the detection of Majorana modes can be performed
by measuring the current flowing through an adjacent quantum dot \cite{Deng2016Dec}.
It turns out that the presence of Majorana zero-energy modes results in unique
transport properties, including fractional values of the conductance $G$
\cite{Liu2011Nov,Leijnse2011Oct,Cao2012Sep,Gong2014Jun,Liu2015Feb,
	Weymann2017Apr,Liu2017Aug,Prada2017Aug,Ptok2017Nov,Gorski2018Oct,Stenger2018Aug,Cifuentes2019Aug,Vernek2019}.
It has been shown that Majorana quasiparticles leaking into the neighboring
dot weakly coupled to external contacts
give rise to $G = (1/4)G_0$, where $G_0=2e^2/h$ \cite{Vernek2014Apr,Ruiz-Tijerina2015Mar}.
On the other hand, in the strong coupling regime, the Majorana-Kondo interplay
determines the transport behavior of the system
\cite{Golub2011Oct,Lee2013Jun,Cheng2014Sep}.
At low temperatures, the quantum interference with a side-attached
topological superconductor results in $G=(3/4)G_0$ \cite{Lee2013Jun}.
Such fractional values of conductance
reveal a half-fermionic nature of Majorana quasiparticles
and may serve as signatures of the presence of these topologically-protected
states in the system \cite{Aguado2017Oct,Lutchyn2018May}.

\begin{figure}[b!]
	\vspace{-0.3cm}
	\includegraphics[width=0.95\columnwidth]{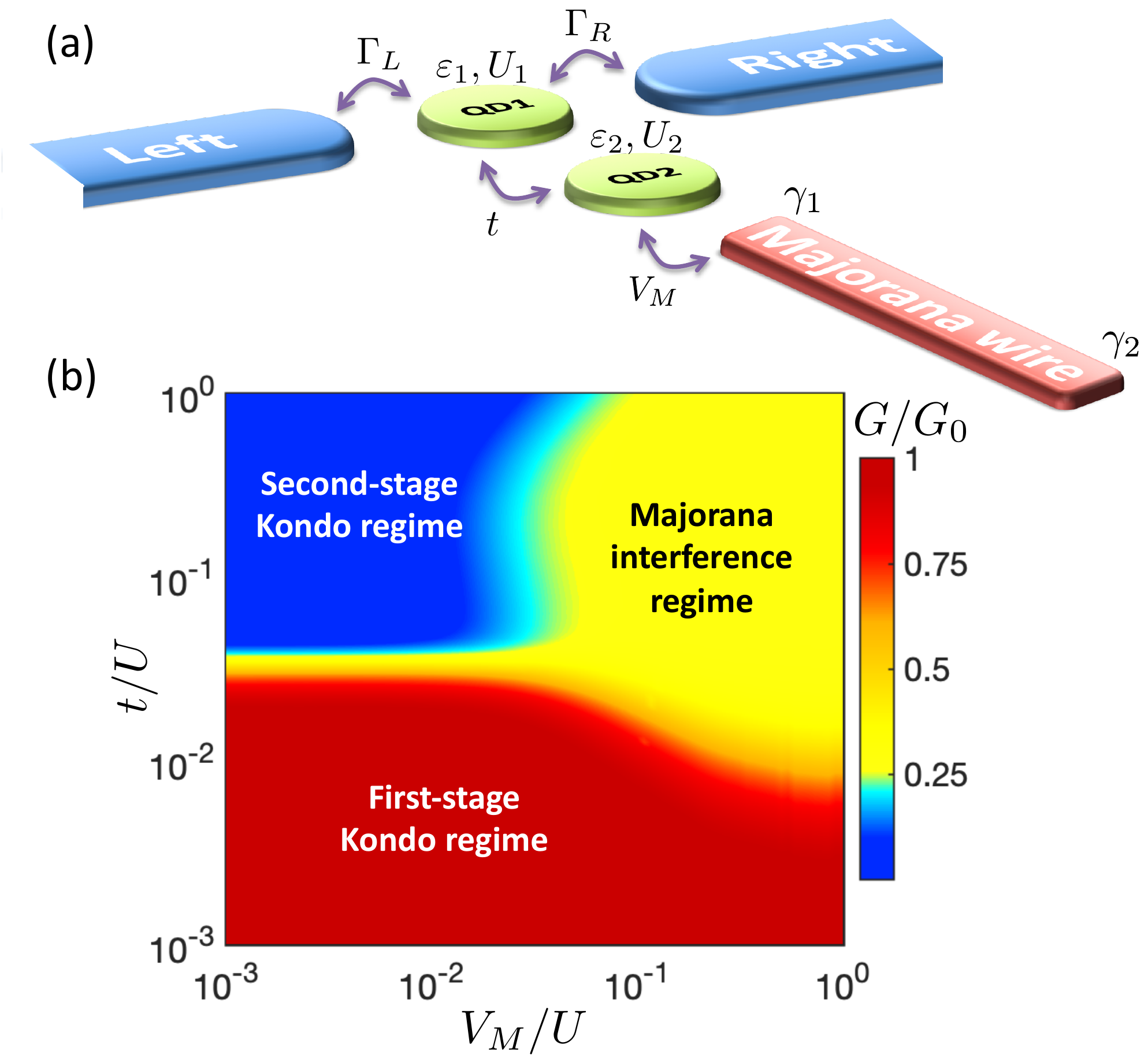}
	\vspace{-0.2cm}	
	\caption{\label{Fig:schematic}
		(a) Schematic of the considered system.
		The first quantum dot is coupled to the left and
		right metallic leads with coupling strengths $\Gamma_L$
		and $\Gamma_R$, respectively.
		The second dot is coupled to the first one through the hopping $t$
		and to the Majorana wire by $V_M$. The wire hosts
		Majorana quasiparticles described by operators $\gamma_1$ and $\gamma_2$.
	    (b) The linear conductance as a function of $V_M$ and $t$ calculated at temperature
    $T=0.0005U$. For parameters see the main text and \fig{Fig:2}.
    Note that $G$ changes monotonically between different transport regimes.
    }
\end{figure}

In this paper we advance further the investigations of interplay
between the Majorana zero-energy modes and correlations giving rise to the Kondo physics
\cite{Kondo1964,Hewson1997,Goldhaber1998},
focusing on the system built of a double quantum dot 
side-coupled to a Majorana wire, as schematically shown in \fig{Fig:schematic}(a).
The double dot is assumed to form with external contacts
a T-shaped geometry, where only one of the dots
is directly coupled to the leads, while the second dot is attached
indirectly through the first quantum dot. This system, in the 
absence of Majorana wire, is known to exhibit
a nonmonotonic dependence of conductance when lowering temperature
due to the two-stage Kondo effect
\cite{Pustilnik2001Nov,Vojta2002Apr,Cornaglia2005Feb,Zitko2006Jan,Chung2008Jan,Sasaki2009Dec}.
Moreover, the interplay between the Fano and Kondo effects in such systems was shown 
to give rise to interesting spin-resolved transport behavior
\cite{DiasdaSilva2013May,Wojcik2014,Wojcik2015Apr}.

The main goal of this work is to uncover unique transport features resulting
from the presence of Majorana quasiparticles
and, in particular, to understand their influence on the two-stage Kondo effect.
To achieve this goal in the most accurate way, 
we make use of the numerical renormalization group (NRG) method \cite{Wilson1975,Bulla2008}.
We study the behavior of the spectral functions
and the temperature dependence of the linear conductance,
which reveal local extrema signaling the leakage of 
Majorana quasiparticles into the double dot system.
Interestingly, we find that the quantum interference with Majorana zero-energy mode
half-suppresses the spectral features related to the second-stage Kondo effect,
giving rise to the fractional value of the linear conductance $G=(1/4)G_0$.
This finding is presented in \fig{Fig:schematic}(b),
which displays the conductance as a function of coupling
to Majorana wire $V_M$ and the hopping between the dots $t$
plotted on logarithmic scale.
One can clearly identify three different transport regimes.
When the first-stage Kondo effect dominates $G=G_0$,
on the other hand, when the system is in the second-stage Kondo regime
$G$ becomes suppressed and reaches $G=0$. However,
once the quantum interference with the Majorana wire becomes relevant
the conductance is given by $G = (1/4)G_0$.
This change from $G=0$ to $G=(1/4)G_0$ is a huge relative difference,
giving better hope for an experimental observation as compared to the reduction
of conductance from $G=G_0$ to $G=(3/4)G_0$, which is present in the case of a single 
quantum dot variant of the system.

Moreover, we show that due to the presence of Majorana quasiparticles
the spectral function exhibits a unique 
five-peak structure, when the device is in the two-stage Kondo regime.
We also demonstrate that, contrary to the expectations based
on the analysis of excitation spectrum of double dot decoupled from normal contacts,
increasing the coupling to topological superconductor
actually enhances the second-stage Kondo temperature. 
A somewhat similar effect has been predicted for single quantum dots
attached to Majorana wires, where an enhancement of the conventional Kondo temperature $T_K$
was observed \cite{Lee2013Jun,Ruiz-Tijerina2015Mar,Wojcik2017,Gorski2018Oct}.
In the T-shaped double quantum dot setup, at $T<T_K$,
the spin of the second dot becomes screened by Fermi liquid
formed by many-body Kondo state generated at the first quantum dot.
Increasing the coupling to Majorana wire results then
in an enhancement of the second-stage Kondo temperature $T^*$,
similarly to the single quantum dot case
where $T_K$ increases with $V_M$ \cite{Lee2013Jun,Ruiz-Tijerina2015Mar,Wojcik2017,Gorski2018Oct}.
At this point we would like to note that transport properties of 
a similar double dot system have been recently studied 
in the regime of relatively large inter-dot hopping \cite{Cifuentes2019Aug}. 
However, the interplay of Majorana quasiparticles
with the correlations giving rise to the two-stage Kondo screening has not been addressed so far,
yet it manifest itself in spectral functions of all the parts of the nanostructure,
as explained in Sec.~\ref{Sec:4A}.

The paper is structured as follows. The model, method and
quantities of interested are presented in Sec.~\ref{sec:model}.
Then, Sec.~\ref{sec:analytics} is devoted to the analysis of eigenspectrum
of the effective Hamiltonian and the discussion 
of the effective exchange interaction between the dots.
Sec.~\ref{sec:results} contains the main results of the paper and their discussion.
Finally, the paper is summarized in Sec.~\ref{sec:summary}.

\section{Theoretical description}
\label{sec:model}

The considered system consists of a double quantum dot in a T-shaped geometry,
i.e. with only one quantum dot attached directly to the leads and the second dot
coupled to the first one through the corresponding hopping matrix elements.
Additionally, the second quantum dot is coupled to a topological superconducting
wire hosting Majorana zero-energy modes at its ends (Majorana wire).
The schematic illustration of this system is presented in \fig{Fig:schematic}(a).
The studied system can be described by the following Hamiltonian
\be \label{Eq:H}
H = H_{\rm Leads} + H_{\rm Tun} + H_{\rm DDM}.
\ee
Here, the first term models the left ($r=L$) and right ($r=R$)
metallic leads as reservoirs of noninteracting quasiparticles
\be
H_{\rm Leads} = \sum_{r=L,R}\sum_{\mathbf{k}\sigma}
\e_{r\mathbf{k}} c^\dag_{r\mathbf{k}\sigma} c_{r\mathbf{k}\sigma},
\ee
where $c_{r\mathbf{k}\sigma}^\dag$ is the creation operator
for an electron with spin $\sigma$, momentum
$\mathbf{k}$ and energy $\e_{r\mathbf{k}}$ in the lead $r$.
The second term of $H$ accounts for tunneling processes between
the double quantum dot-Majorana wire subsystem and the normal
leads. Because in the considered setup only the first dot is directly coupled
to electrodes, the tunneling Hamiltonian simply reads
\be
H_{\rm Tun} = \sum_{r=L,R}\sum_{\mathbf{k}\sigma} v_{r} \left(d^\dag_{1\s}
c_{r\mathbf{k}\sigma} + c^\dag_{r\mathbf{k}\sigma} d_{1\s} \right),
\ee
with the corresponding tunnel matrix elements described by $v_r$ and
assumed to be momentum independent.
The operator $d^\dag_{1\s}$ ($d_{1\s}$) is the
creation (annihilation) operator
of an electron with spin $\s$ in the first quantum dot.
The coupling to external leads gives rise to the broadening
of the first dot level, which can be described by
$\Gamma_r = \pi \rho_{r} v_{r}^2$, where
$\rho_{r}$ is the density of states of a given lead.
In these considerations we assume a flat band of width
$2D$ for each electrode and take $\Gamma_L = \Gamma_R \equiv \Gamma/2$.
The band halfwidth is hereafter used as the energy unit $D\equiv 1$.

Finally, the last term of the Hamiltonian $H$ models the double dot-Majorana wire
subsystem, and it can be written as
\bea \label{Eq:HDDM}
H_{\rm DDM} &=& \sum_{j=1,2}\sum_{\s} \e_j d_{j\s}^\dag d_{j\s}
+ \sum_{j=1,2} U_j d_{j\uparrow}^\dag d_{j\uparrow} d_{j\downarrow}^\dag d_{j\downarrow}
\nonumber\\
&&+ \sum_\s t (d_{1\s}^\dag d_{2\s} +  d_{2\s}^\dag d_{1\s})
\nonumber\\
&&+ \sqrt{2} V_M (d^\dag_{2\downarrow} \gamma_1 + \gamma_1 d_{2\downarrow}) 
+ i \e_M \gamma_1 \gamma_2.
\eea
Here, $d_{j\s}^\dag$ is the creation operator
for a spin-$\s$ electron on dot $j$ with the energy $\e_j$,
and the two electrons residing on the same dot interact
with the Coulomb correlation energy $U_j$.
For the sake of clarity and convenience,
we assume $U_1=U_2\equiv U$.
The two quantum dots are coupled through
the hopping matrix elements $t$.
The coupling to the Majorana wire is 
described by the penultimate term of $H_{\rm DDM}$,
where $V_M$ is the corresponding tunneling matrix element
\cite{Flensberg2010Nov,Liu2011Nov,Lee2013Jun,Weymann2017Apr,Hoffman2017Jul}.
The Majorana quasiparticles localized at the ends
of the topological superconductor wire are described by 
the operators $\gamma_1$ and $\gamma_2$.
The overlap between the wave functions of these two quasiparticles
is described by $\e_M$.
When the length of the Majorana wire is much larger
than the superconducting coherence length,
the two Majorana quasiparticles do not overlap
and, consequently, $\e_M = 0$ \cite{Albrecht2016Mar}.
In the opposite case, $\e_M$ is finite, which results in a splitting
of the energies of the Majorana quasiparticles.
In the following we will refer to these two situations
as the case of long/short Majorana wire.

The Majorana operators $\gamma_1$ and $\gamma_2$
can be represented by a fermionic operator $f$
as ${\gamma_1 = (f^\dag+f)/\sqrt{2}}$
and ${\gamma_2 = i(f^\dag-f)/\sqrt{2}}$, respectively.
Then, the last two terms of $H_{\rm DDM}$ can be expressed as
\bea \label{Eq:VM}
{\sqrt{2} V_M (d^\dag_{2\downarrow} \gamma_1 + \gamma_1 d_{2\downarrow})} &=& 
{V_M (d^\dag_{2\downarrow}-d_{2\downarrow}) (f^\dag+f)}, \label{feq1}\\
i \e_M \gamma_1 \gamma_2 &=& \e_M (f^\dag f - 1/2). \label{feq2}
\eea

We note that since the Hamiltonian of the double dot coupled to normal leads possesses
the full spin $SU(2)$ symmetry, one can choose the quantization axis in such a way
that only one of the spin components couples to the Majorana mode
\cite{Flensberg2010Nov,Liu2011Nov,Lee2013Jun}.
In our considerations, we assumed that the spin-down component
is coupled to Majorana quasiparticles, cf. Eq.~\eqref{Eq:HDDM}.
However, to make the analysis more general, in Sec.~\ref{sec:2spins} we also present the results for the case when
the Majorana zero-energy modes are coupled to both spin projections \cite{Hoffman2017Jul}.

In this paper we are mainly interested in the linear response transport properties
of the considered Majorana-double dot structure.
The linear conductance between the left and right contacts can be then found from \cite{Meir1992Apr}
\be\label{Eq:G}
G = \frac{e^2}{h} \sum_\s \int \! d\w \left[ -f'(\w) \right] \pi\G A_\s (\w),
\ee
where $f'(\w)$ is the derivative of the Fermi-Dirac distribution function.
$A_\s (\w)$ denotes the spectral function of the first quantum dot
for spin $\sigma$, $ A_\s (\w) = -\tfrac{1}{\pi} {\rm Im } G^R_\s(\w)$,
where $G^R_\s(\w)\equiv \GFzub{d_{1\s}}{d_{1\s}^\dag}_\omega^R$ is the Fourier transform of the retarded
Green's function $G^R_\s(t) = -i\Theta(t) \GF{d_{1\s}(t)}{d_{1\s}^\dag(0)} $.
To obtain the most reliable results and quantitatively understand the interplay of 
strong electron correlations with the presence of Majorana zero-energy modes,
we use the numerical renormalization group method \cite{Wilson1975,Bulla2008,NRG_code}.
In NRG calculations we use the discretization parameter
$\Lambda=2$ and keep at least $3000$ states at each iteration.
Moreover, to increase the accuracy of the spectral functions,
which are typically subject to broadening issues \cite{Zitko2009Feb},
we average the data over $4$ discretizations \cite{Campo2005Sep}
and use the optimal broadening method \cite{Freyn2009Mar}.
On the other hand, the results presented for the linear conductance are
obtained directly from the discrete NRG data,
without the need of resorting to broadening \cite{Weymann2013Aug}. 

\section{Effective exchange interaction}
\label{sec:analytics}

In general, the low-temperature transport behavior of a system 
depends mostly on the low-energy part of the spectrum.
In the T-shaped double quantum dot in the presence of normal leads
the low-energy states {\it relevant} for the two-stage Kondo regime are those consisting of two 
singly-occupied quantum dots, organized into singlet and triplet,
split by the effective antiferromagnetic exchange interaction $\Jeff\approx 4t^2/U$ 
\cite{Cornaglia2005Feb}. This structure remains untouched when the device is 
proximized by the conventional superconductor, only the value of $\Jeff$
increases (irrespective of the geometry), or even $\Jeff$ can arise due to the coupling 
to a BCS-like superconductor due to crossed Andreev reflection processes \cite{KWIW-2SQD_2019}.
However, the situation qualitatively changes when the superconductor is topological.

To explore such a case, we define the basis states as $\ket{\chi_1 \chi_2 n_f}$,
where $\chi_1$ and $\chi_2$ are the local states of quantum dots $1$ and $2$, with 
$\chi_i \in \{0,\up,\down,2\}$, and the Majorana zero-energy modes are described
by the occupation of the auxiliary fermionic operator $f$, $n_f \in \{0, 1\}$;
cf. Eqs. (\ref{feq1}) and (\ref{feq2}).
The local Hamiltonian consists then of $32$ states, nevertheless,
the states relevant for the Kondo regime still consist of half-filled 
quantum dots. There are $8$ such states and we will refer to them
as {\it relevant} states henceforth.

\begin{table*}[t!]
	\begin{tabular}{c @{$\;+\;$} c @{$\;+\;$} c @{$\;+\;$} c @{\hspace{4ex}} c @{\hspace{2ex}} c @{\hspace{2ex}} c @{\hspace{2ex}} c}
		\multicolumn{4}{c}{State} 					&	$Q_P$	&	$I$	&	$I_z$ & 	Energy	\\
		\hline
		$\alpha_1 \ket{{\rm s},\!1} $
         &$\beta_{11}	(\ket{\downx 20}\!+\!\ket{\upx 00})/\sqrt{2}$
         &$\beta_{12}	(\ket{2\downx 0}\!+\!\ket{0\upx 0})/\sqrt{2}$
         &$\beta_{13}	(\ket{021}\!+\!\ket{201})/\sqrt{2}  $	
		&	$-1$ & $0$ & $0$ &	$E^0_{++}$\\
		\hline
		$\alpha_1 \ket{{\rm s},\!0} $
         &$\beta_{11}	(\ket{\downx 21}\!+\!\ket{\upx 01})/\sqrt{2}$
         &$\beta_{12}	(\ket{2\downx 1}\!+\!\ket{0\upx 1})/\sqrt{2}$
         &$\beta_{13}	(\ket{020}\!+\!\ket{200})/\sqrt{2}  $
		&	$+1$ & $0$ & $0$ &	$E^0_{++}$\\
		\hline
		$\alpha_2 \ket{\down\down 1} $
		& $\beta_{21}\ket{\downx 00}$
		& $\beta_{22}\ket{0 \downx 0}$
		& $\beta_{23}\ket{001} $ 
		&	$-1$ & $1$ & $-1$ & $E^1_{++}$\\
		$\alpha_3 \ket{{\rm t},\!1} $
		& $\beta_{31}(\ket{\downx 20}\!-\!\ket{\upx 00})/\sqrt{2}$
		& $\beta_{32}(\ket{0\upx 0}\!+\!\ket{2\downx 0})/\sqrt{2}$
		& $\beta_{33}(\ket{201}\!-\!\ket{021}) /\sqrt{2}$	
		&	$-1$ & $1$ & $0$ & $E^1_{++}$\\
		$\alpha_4 \ket{\up\up 1} $
		& $\beta_{41}\ket{\upx 20}$
		& $\beta_{42}\ket{2 \upx 0}$
		& $\beta_{43}\ket{221}$ 	
		& 	$-1$ & $1$ & $+1$ & $E^1_{++}$\\
		\hline
		$\alpha_2 \ket{\down\down 0} $
		& $\beta_{21}\ket{\downx 01}$
		& $\beta_{22}\ket{0 \downx 1}$
		& $\beta_{23}\ket{000} $ 
		&	$-1$ & $1$ & $-1$ & $E^1_{++}$\\
		$\alpha_3 \ket{{\rm t},\!0} $
		& $\beta_{31}(\ket{\downx 21}\!-\!\ket{\upx 01})/\sqrt{2}$
		& $\beta_{32}(\ket{0\upx 1}\!+\!\ket{2\downx 1})/\sqrt{2}$
		& $\beta_{33}(\ket{200}\!-\!\ket{020}) /\sqrt{2}$	
		&	$-1$ & $1$ & $0$ & $E^1_{++}$\\
		$\alpha_4 \ket{\up\up 0} $
		& $\beta_{41}\ket{\upx 21}$
		& $\beta_{42}\ket{2 \upx 1}$
		& $\beta_{43}\ket{220}$ 	
		& 	$-1$ & $1$ & $+1$ & $E^1_{++}$\\
		\hline
	\end{tabular}
	\caption{The eight lowest-energy eigenstates of the local Hamiltonian, \eq{Eq:HDDM}, 
		obtained for $\e_1=\e_2=-U/2$, $\e_M=0$ and $U_1=U_2=U$. For brevity the following notation
		for the singlet and triplet states,
		$\ket{{\rm s},n_f} = (\ket{\down\up n_f} - \ket{\up\down n_f})/\sqrt{2}$ and 
		$\ket{{\rm t},n_f} = (\ket{\down\up n_f} + \ket{\up\down n_f})/\sqrt{2}$, was used.
		The first column presents the corresponding eigenstate, with $\alpha_{a}$ and
		$\beta_{ab}$ being the coefficients of the corresponding states. 
		$I$ and $I_z$ stand for the isospin quantum numbers.
		Note that for $t,V_M \ll U$ the coefficient of the {\it relevant} state dominates, 
		$|\alpha_a| \sim 1$. Moreover, after neglecting $\beta_{ab}$ the isospin $I$ becomes 
		equivalent to the physical spin $\S$.}
	\label{tabExact}
\end{table*}

Furthermore, since in the considered model the Majorana mode couples to one spin channel,
the spin $\S$ is no longer a good quantum number, nor are its components.
Thus, the structure of the eigenbasis cannot be determined from the spin symmetry
requirements. However, the system still exhibits two symmetries, namely, related to the conservation 
of fermion number {\it parity}, $Q_P = (-1)^Q$ [with 
$Q = \fk f + \sum_{j\s}\dk_{j\s}d^{}_{j\s}$ 
denoting the fermion number] and the conservation of the number of spin-up electrons, 
$N_{\up} = \sum_j \dk_{j\up}d^{}_{j\up}$ \cite{Bulla1997Jun,Bradley1999Oct}. 
Moreover, for half-filled quantum dots (obtained in the model by setting $\e_1=\e_2=-U/2$)
and long Majorana wire ($\e_M=0$),  the Abelian symmetry related to the $N_\up$ conservation 
is generalized to the full ${\rm SU}(2)$ isospin symmetry
with its $z$ component defined as 
\begin{equation}
I_z = \sum_j (d^\dagger_{j\uparrow}d_{j\uparrow} - \tfrac{1}{2}).
\label{Iz}
\end{equation}
The operators rising and lowering $I_z$ can be defined as
\begin{equation}
I_+ = \sum_j \dk_{j\up} \left[ d^{}_{j\down} + (-1)^{j} \dk_{j\down} \right],
\label{Ip}
\end{equation}
%
and $I_- = (I_+)^\dagger$.
Note that while the on-site "spin-flip" part of $I_+$ (proportional to $\dk_{j\up}d^{}_{j\down}$) has
always the same sign, the "charge-flip" term has an alternating sign, analogously to the 
charge ${\rm SU}(2)$ symmetry generators. Importantly, the symmetry naturally extends to the full 
Wilson chain representation of the Hamiltonian (\ref{Eq:H}) used in NRG calculations.
This is done by allowing $j$ to run in the range $[-2,N]$ (with chain sites numbered from $0$ to $N$),
where $j=-2$ and $j=-1$ correspond to the second and first quantum dot,
which can be incorporated into the chain, 
and identifying $d_{-j\s}$ with the operator 
$f_{j\s}$ acting at the $j$-th site of the Wilson chain.

\begin{figure}[b]
	\includegraphics[width=0.9\columnwidth]{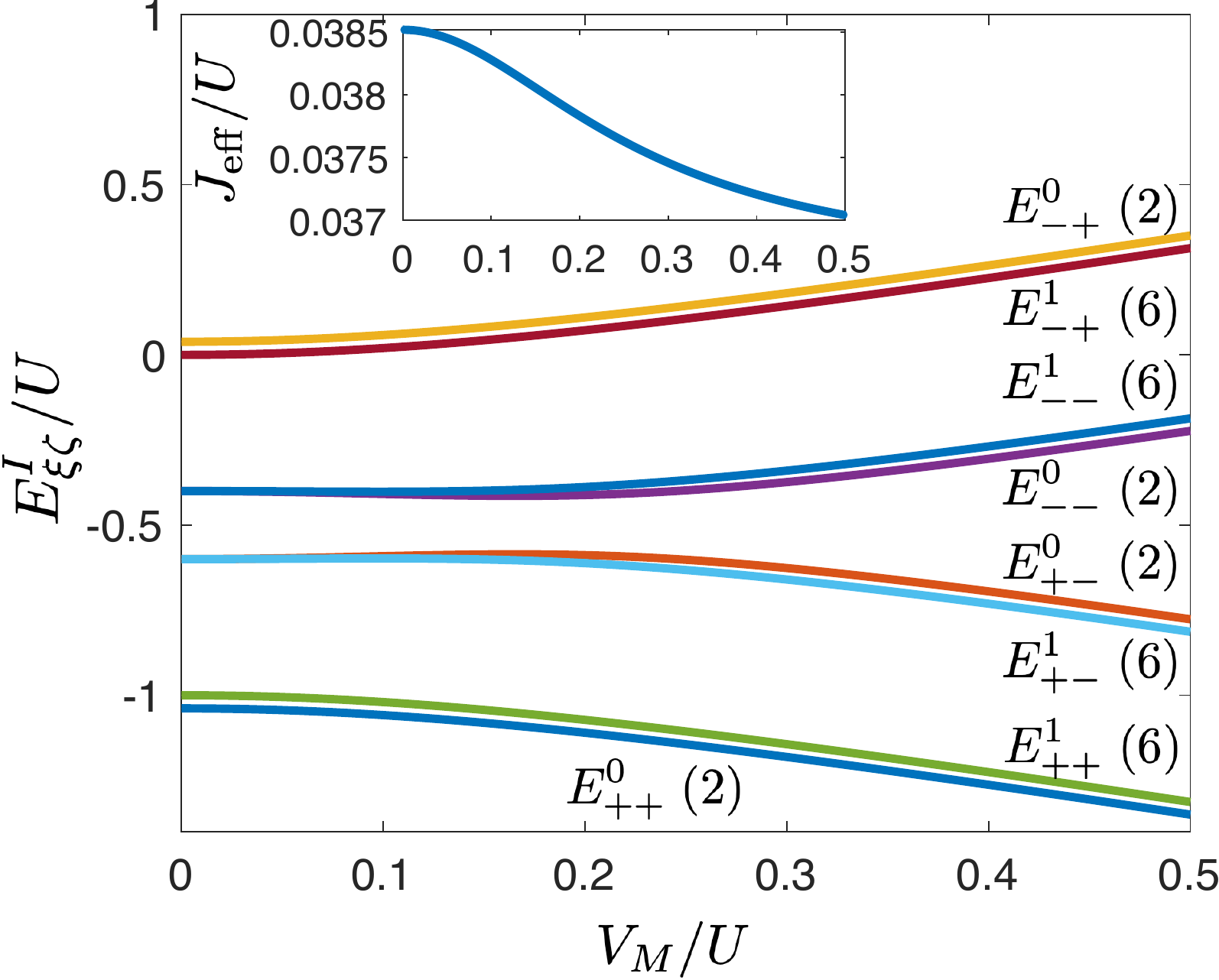}
	\caption{\label{Fig:E}
		The eigenspectrum of the double dot-Majorana wire effective
		Hamiltonian, Eq.~(\ref{Eq:HDDM}), calculated as a function of $V_M$
		for $\e_1=\e_2=-U/2$, $t=0.1 U$ and $\e_M=0$.
		The corresponding energies given by \eq{E01} are indicated,
		together with the degeneracy of states (numbers in brackets).
		The inset presents the effective exchange interaction
        $\Jeff \equiv E^{1}_{++} - E^{0}_{++}$.		
	}
\end{figure}

The isospin symmetry defined above can be easily
recognized in the spectrum of the local Hamiltonian,
$H_{\rm DDM}$. Its eigenenergies have the form 
\begin{equation}
E^{I}_{\xi\zeta} = -\frac{U}{2} \left( 1 + \frac{\xi}{\sqrt{2}}\sqrt{A_I+\zeta \sqrt{B_I}} \right),
\label{E01}
\end{equation}
where indices $\xi, \zeta$ take values $\pm 1$ and
\begin{eqnarray*}
A_1 &=& 1 + 4 (t^2+2V_M^2) U^{-2} ,
	\label{A1}\\
B_1 &=& 1 - 8 (t^2 -2V_M^2) U^{-2} + 16 t^2 (t^2+4V_M^2) U^{-4},
	\label{B1}\\
A_0 &=& A_1 + 16 t^2 U^{-2},
	\label{A0}\\
B_0 &=& B_1 + 32 t^2 U^{-2} + 128 t^2 (t^2+4V_M^2) U^{-4} ,
	\label{B0}
\end{eqnarray*}
are constants of the order of unity.
The four combinations of signs $(\xi,\zeta)$,
together with four combinations of isospin and its $z$-component $(I,I_z)$ 
corresponding to $I\leq 1$, and double degeneracy due to $Q_P$, give all the $32$ local states. 
The eigenenergies are plotted as a function of $V_M$,
together with their degeneracy, are shown in Fig.~\ref{Fig:E}.

For $U \gg t,V_M$, the order of magnitude of the energies given by \eq{E01} is 
determined by the signs of $\xi$ and $\zeta$. The lowest energies correspond to $\xi=+1$ and $\zeta=+1$, 
then $E^I_{++}\sim -U$; the other options lead to $E^I_{\xi\zeta}\sim -U/2$ or $E^I_{\xi\zeta} \sim 0$---up to
the terms of the order of $U^{-2}$. For the analysis of the low-temperature properties only the former of these 
are important. Those states, together with their quantum numbers,  are listed in Table~\ref{tabExact}.
Note, that even though in general the eigenstates do not possess a definite spin $\S$,
when projected onto the subspace spanned by the {\it relevant} states
(when the coefficients $\beta_{ab}$, defined in Table~\ref{tabExact},
are neglected) they actually do, {\it i.e.} 
$I$ multiplets correspond to $\S$ multiplets with $I=\S$. In general, however,
each of the $(I,I_z)$ eigenstates is a superposition of a single relevant state with $(\S,\S_z)=(I,I_z)$
and a number of other states, as presented in Table~\ref{tabExact}.

The relation between $\S$ and $I$ allows us actually  to define the effective exchange interaction
as the difference between the energies of the low-energy isospin singlet and triplet states,
\begin{eqnarray}
\Jeff &\equiv& 	E^{1}_{++} - E^{0}_{++} \nn\\
		& = & 	\frac{U}{2\sqrt{2}} \left( \sqrt{A_0+\sqrt{B_0}} - \sqrt{A_1+\sqrt{B_1}}\right)  \nn\\
		& \approx &	\frac{4 t^2}{U} \left[1 - \left(\frac{2 t}{U}\right)^{\!\! 2}\! + 2 \left(\frac{2 t}{U}\right)^{\!\! 4}\!
		 - 5 \left(\frac{2 t}{U}\right)^{\!\! 2}\! \left(\frac{2 V_M}{U}\right)^{\!\! 2} \right]
		\nn\\ &&+ \mathcal{O}(U^{-7}).
\label{Jeff}
\end{eqnarray}
Note that the effect of coupling to topological superconductor becomes relevant only in the 
fifth order of expansion with respect to $1/U$.
Clearly, from \eq{Jeff} one can conclude that the coupling to the Majorana wire slightly decreases the effective 
exchange interaction between the quantum dots.
This can be explicitly seen in the inset to Fig.~\ref{Fig:E},
which presents the dependence of 
$\Jeff  = E^{1}_{++} - E^{0}_{++}$ on $V_M$.
Despite very small magnitude of the decrease of $J_{\rm eff}$, 
one could expected a noticeable decrease of $T^*$, due to its 
exponential dependence on the inter-dot exchange; cf. \eq{Tstar}.
Interestingly, this effect is opposite to what happens in the 
presence of a conventional superconductor \cite{KWIW-2SQD_2018,KWIW-2SQD_2019}.
However, as shown in the following by accurate NRG calculations,
the decrease of {\it bare} exchange interaction becomes overwhelmed by strong electron correlations.
Actually, we demonstrate that increasing the coupling to topological wire
results in an {\it enhancement} of the second-stage Kondo temperature $T^*$
instead of reduction, as one could expect from simple analysis of $H_{\rm DDM}$ spectrum, cf. Eq. (\ref{Jeff}).

We note that a similar effect has been predicted for
single quantum dots coupled to normal leads and a topological superconductor,
where increasing the coupling to Majorana wire gives rise to an enhancement 
of the Kondo temperature \cite{Lee2013Jun,Ruiz-Tijerina2015Mar,Wojcik2017,Gorski2018Oct}.
In the setup considered in this paper, at energy scales below the first-stage Kondo temperature $T_K$,
the double dot system can be viewed as an effective single quantum dot attached to
a conduction band of width $T_K$ (resulting from Fermi liquid formed by first quantum dot
screened by lead conduction electrons) and additionally coupled to Majorana wire.
Then, one could expect that increasing the coupling to topological wire
would result in an increase of the relevant Kondo temperature (the second-stage Kondo temperature $T^*$),
similarly as it does in the case of single quantum dots 
\cite{Lee2013Jun,Ruiz-Tijerina2015Mar,Wojcik2017,Gorski2018Oct}.
This picture wins over the local-Hamiltonian perspective presented in this section, as is shown
by NRG calculations presented in the following.

\section{Numerical results and discussion}
\label{sec:results}

We now turn to the numerical analysis
of the transport behavior of the considered system.
First, we consider the case of long Majorana wire and then
also discuss the situation when there is a finite overlap between 
the Majorana quasiparticles. Finally, at the end, we examine
the case when the Majorana wire is coupled to both 
spin projections of the double dot.
To uncover the interplay between
the Majorana and Kondo physics,
we study the behavior of the relevant spin-resolved  spectral functions
as well as the temperature and gate voltage
dependence of the linear conductance through the system.

To set the background for the following discussion,
let us begin with a short introduction to the case of $V_M=0$.
As already mentioned in the Introduction, in such a situation 
the system exhibits the two-stage Kondo effect,
which is governed by two energy scales, $T_K$ and $T^*$
\cite{Cornaglia2005Feb,Chung2008Jan}.
With lowering the temperature, the Kondo effect develops on the 
first quantum dot once $T\lesssim T_K$, where the first-stage Kondo temperature $T_K$
for $\e_1=-U/2$ can be estimated from \cite{Haldane1978Feb}
\be \label{TK}
T_K \approx \sqrt{\tfrac{\Gamma U}{2}} e^{-\pi U / 8\Gamma}\,.
\ee
When the temperature is decreased further, such that $T\lesssim T^*$,
the spin on the second quantum dot becomes screened by the Fermi liquid formed by 
first dot strongly coupled to the leads. The second-stage 
Kondo temperature $T^*$ can be evaluated from
\cite{Cornaglia2005Feb,Zitko2006Jan,Wojcik2015Apr}
\be
T^* \approx \alpha T_K e^{-\beta T_K/\Jeff},
\label{Tstar}
\ee
where $\alpha$ and $\beta$ are constants of the order of unity
and $\Jeff\approx 4t^2/U$.
This consecutive screening results in a nonmonotonic dependence
of the spectral function on energy, see \fig{Fig:2}(a) for $V_M=0$,
as well as a nonmonotonic temperature dependence of the 
linear conductance; for $G(T)$ in the case of $V_M=0$ see \figs{Fig:4}(a).
When $T$ decreases, the conductance initially increases due to the first-stage Kondo effect,
however, when the second spin experiences screening at even lower temperatures,
it effectively scatters electrons transported through the central quantum dot and
$G$ becomes suppressed.

As can be seen from Eqs. (\ref{TK}) and (\ref{Tstar}),
both temperatures strongly depend on the relevant tunnel matrix elements
and may be thus tuned by gate voltages.
Moreover, because $T^*$ depends exponentially on the effective exchange interaction $\Jeff$
between the two quantum dots generated by the hopping $t$,
changing $t$ results in large changes in $T^*$ [see also \fig{Fig:4}(a)].
We would also like to notice that recently the two-stage Kondo effect has been explored experimentally
down to temperatures much lower than $T^*$ \cite{Guo2020Mar}.

\subsection{Spectral functions}
\label{Sec:4A}

\begin{figure}[t!]
	\includegraphics[width=1\columnwidth]{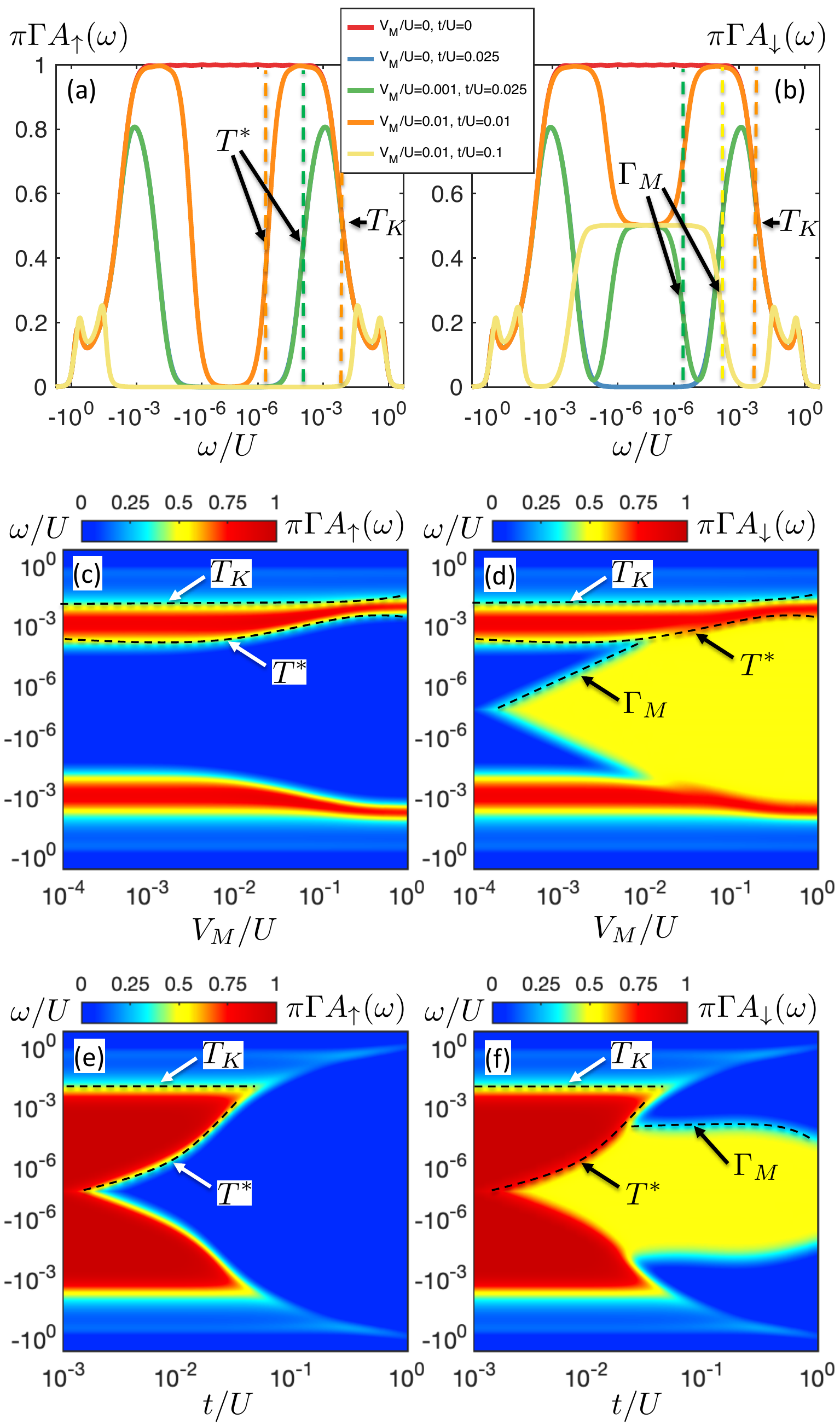}
	\caption{\label{Fig:2}
		The normalized spectral function $\pi\Gamma A_\sigma(\omega)$ of the first quantum dot
		for (left column) spin-up and (right column) spin-down components.
		The first row presents the spectral function
		plotted as a function of energy $\omega$,
		where also the relevant energy scales are marked
		by vertical dashed lines.
		The next two rows display the density plots
		of the spectral function as function of 
		energy and (c,d) the coupling to Majorana wire $V_M$ for $t/U = 0.025$
		and (e,f,) the hopping between the dots for $V_M/U = 0.01$.
		The other parameters are: $U=0.2D$, $\e_1 = \e_2 = -U/2$,
		$\Gamma = 0.1U$, where $D$ is the band halfwidth
		taken as energy unit $D\equiv 1$.
		Note the logarithmic scale on the axes.
		}
\end{figure}

\begin{figure}[t!]
	\includegraphics[width=1\columnwidth]{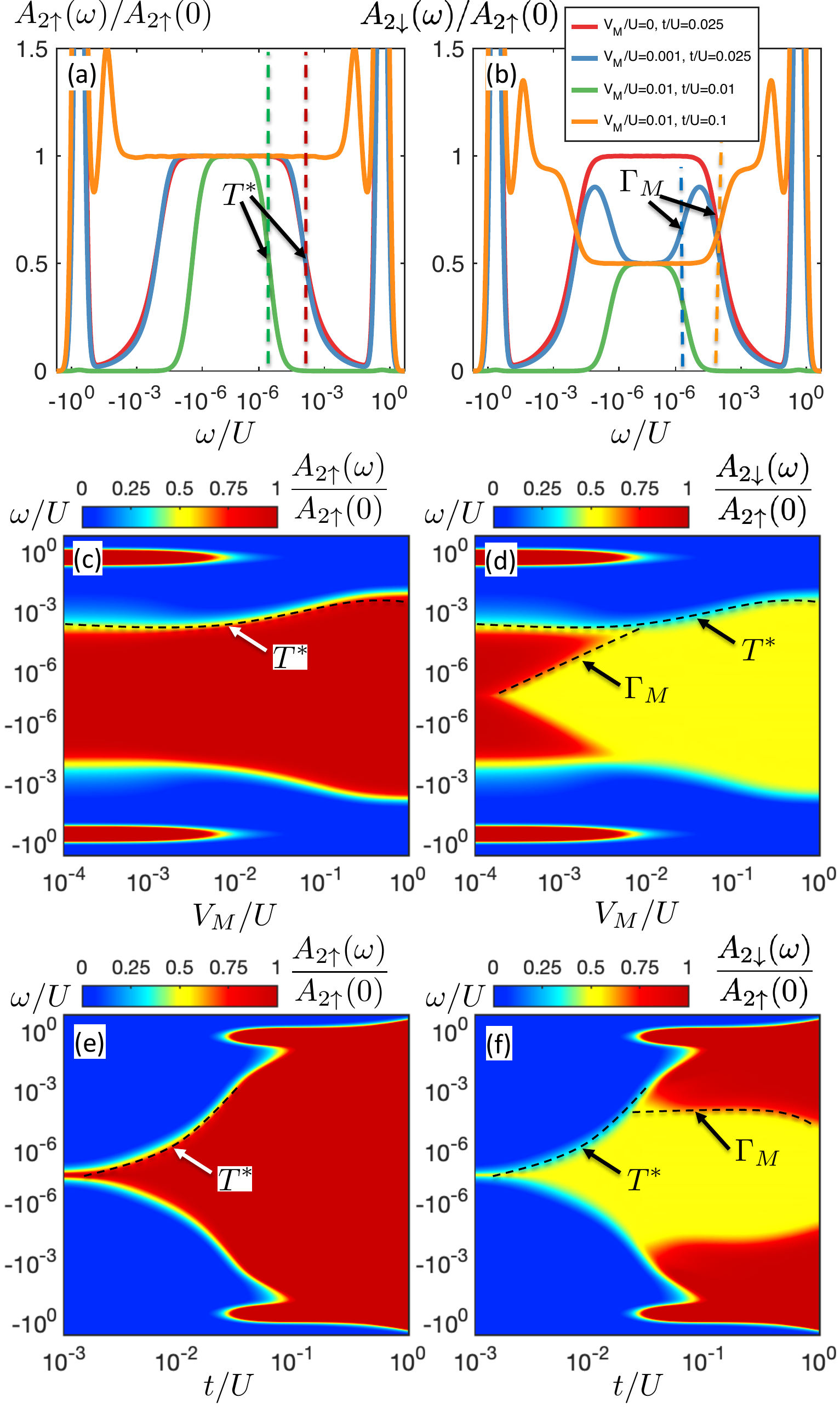}
	\caption{\label{Fig:3}
		The same as in \fig{Fig:2} calculated for
		the spectral function of the second quantum dot
		$A_{2\sigma}(\omega)$. The spectral function
		is normalized to its value at the Fermi energy
		for the spin-up component $A_{2\uparrow}(0)$.
        Note that in order to resolve the low-energy features
        in the density plots
    the color scale has been limited to $1$. However,
the value of the spectral function at resonances $\w=\pm -U/2$
can exceed $A_{2\uparrow}(0)$ depending on $t$, as can be seen in (a) and (b).}
\end{figure}

The spin-resolved spectral function of the quantum dot directly coupled
to the normal leads is shown in \fig{Fig:2}.
The first row shows $A_\sigma(\omega)$
for representative values of the coupling
to Majorana wire $V_M$ and the hopping between the dots $t$.
There are three different energy scales in the system
that determine the transport behavior---these are marked
with vertical dashed lines in \figs{Fig:2}(a) and (b).
In the case of $t=V_M=0$, the system exhibits
the usual spin-$1/2$ Kondo effect \cite{Hewson1997}.
On the other hand, when the hopping between the dots is finite but $V_M=0$,
one observes the conventional two-stage Kondo effect \cite{Cornaglia2005Feb,Chung2008Jan},
see the blue line in the first row of \fig{Fig:2}.
(Note, that this line coincides with the green line for
$V_M/U=0.001$ and $t/U=0.025$ except for low energies
in the spin-down component.)
As can be seen in the figure,
the spectral function first increases with lowering the energy $\omega$,
which happens for $\omega\lesssim T_K$, but then starts to decrease once $\omega\lesssim T^*$.

The behavior of the spectral function changes when the coupling
to Majorana wire is present. Note that in the effective Hamiltonian
we assumed that the spin-down component of the dot's spin
couples to the Majorana quasiparticles. Thus, the largest effects related to the presence
of topological superconductor can be expected
in the behavior of $A_{\down}(\w)$. Nevertheless, 
finite $V_M$, through the Coulomb correlations, also affects
the other spin component of the spectral function.
As can be seen, the spin-up spectral function exhibits the usual two-stage Kondo dependence,
with $A_{\up}(\w\to 0)\to 0$.
This is just opposite to the case of the spin-down spectral function,
where finite values of $V_M$ result in $A_\down(0) = 1/(2\pi\Gamma)$.
Such a fractional value of the spectral function at the Fermi energy
is a direct signature of a half-fermionic nature of Majorana quasiparticles.
The coupling to Majorana wire {\it half}-suppresses the second-stage of
the Kondo effect at a new energy scale $\Gamma_M$
resulting from the coupling to topological superconducting wire.
As a consequence of this suppression,
a five-peak structure can be visible in the spectral function
of the first quantum dot, 
see the curves for $V_M/U=0.001$ and $t/U=0.025$,
and for $V_M/U=0.01$ and $t/U=0.1$ in \fig{Fig:2}(b).
$A_\down(\w)$ exhibits the usual Hubbard resonances 
for $\w\approx \pm U/2$ (note that $\e_1=\e_2=-U/2$).
Then, with lowering $\w$,
$A_\down(\w)$ starts growing 
due to the Kondo effect, however, it becomes suppressed
once $\w \approx  T^*$ due to the second-stage Kondo screening,
which results in a local maximum around $\w \approx T_K$.
With further decrease of $\w$, the Majorana energy
scale $\Gamma_M$ comes into play, destroying 
the second-stage Kondo effect and resulting 
in a further resonance just at the Fermi energy.
This happens when the coupling to the Majorana wire is smaller
than the hopping $t$.
On the other hand, when the coupling to Majorana wire 
is comparable to the hopping, 
see the case for $V_M/U=t/U=0.01$ in \figs{Fig:2}(a) and (b),
both spin resolved spectral functions exhibit a four-peak structure
due to the two-stage Kondo effect. However, the spin-down
component does not drop to zero at the Fermi energy
such as its spin-up counterpart, but retains finite value of $A_\down(0) = 1/(2\pi\Gamma)$.

In turn, we analyze how the relevant energy scales
change with both $V_M$ and $t$. The second row of \fig{Fig:2}
presents the density plots of the spectral function
versus $V_M$ calculated for $t/U=0.025$.
For the spin-up component, it is clearly evident that
the coupling to Majorana wire strongly
affects the second stage Kondo temperature $T^*$,
i.e. $T^*$ grows with increasing $V_M$.
It is also interesting to note that,
although $T^*$ strongly depends on $V_M$, 
the behavior of $A_\up(\omega\to 0)$
does not depend on the coupling to the Majorana wire
and one has $A_\up(\omega\to 0)\to 0$.
In the case of $A_{\down}(\w)$, one observes
that for relatively small values of the coupling to Majorana wire,
the low-energy behavior of the spectral function starts changing.
A plateau of $A_\down(0) = 1/(2\pi\Gamma)$ develops at low energies 
once the energy scale becomes smaller than $\Gamma_M$, see \fig{Fig:2}(d).
With increasing $V_M$ further, $\Gamma_M$ merges with $T^*$
and the characteristic five-peak structure disappears.
Then, an enhancement of $T^*$ with increasing $V_M$ can be observed.

The dependence of $A_{\sigma}(\omega)$ on $t$ for $V_M/U=0.01$
is presented in \figs{Fig:2}(e) and (f).
Since $T^*$ depends strongly on $t$,
the region of $A_\uparrow(\omega)\approx 1/\pi\Gamma $
for $T^*\lesssim T\lesssim T_K$ shrinks as $t$ grows and, e.g. for $t/U = \Gamma/U=0.1$,
the spectral function displays only a small resonance, see Fig. \ref{Fig:2}(e).
This is characteristic of the \textit{local singet} regime,
where the Kondo effect on the first quantum dot does not develop,
but the two dots form a molecular singlet state.
The two-stage Kondo regime can be reached from this phase by reducing the hopping $t$.
We note that even though this is a continuous crossover,
it is related to switching on or off many-body Kondo correlations.
Namely, for large $t$ the Kondo effect is absent, while for small values of the hopping
between the dots the many-body Kondo state develops.
Interestingly, when $T^*$ becomes larger than the Majorana energy scale $\Gamma_M$,
an additional resonance at the Fermi energy of halfwidth $\sim \Gamma_M$
forms in the spin-down spectral function.

To make the discussion more comprehensive, 
we now analyze the behavior of the spectral function
of the second quantum dot $A_{2\sigma}(\omega)$,
which is presented in \fig{Fig:3}. This figure is calculated
for the same parameters as \fig{Fig:2} and presents
the same dependencies. The spectral function
is normalized to the value of the spin-up component taken
at the Fermi energy $A_{2\up}(0)$.
When the system is in the two-stage Kondo regime,
the spin-up spectral function displays a plateau at low energies
of halfwidth $\sim T^*$. On the other hand,
the spin-down spectral function also exhibits a plateau
of the same width, but half-reduced magnitude, i.e.
$A_{2\down}(0) = A_{2\up}(0)/2$. This is the signature of the
presence of Majorana zero-energy mode and its half-fermionic nature.
The density plots of $A_{2\sigma}(\omega)$ presented in \fig{Fig:3}
clearly reveal the behavior of the relevant energy scales,
which is similar to that shown in \fig{Fig:2}.
One can conclude, that the signatures of the Majorana physics
are visible in the spectra of both quantum dots, and it is not 
justified to prescribe the presence of a Majorana quasiparticle to any of them.

\begin{figure}[t!]
	\includegraphics[width=1\columnwidth]{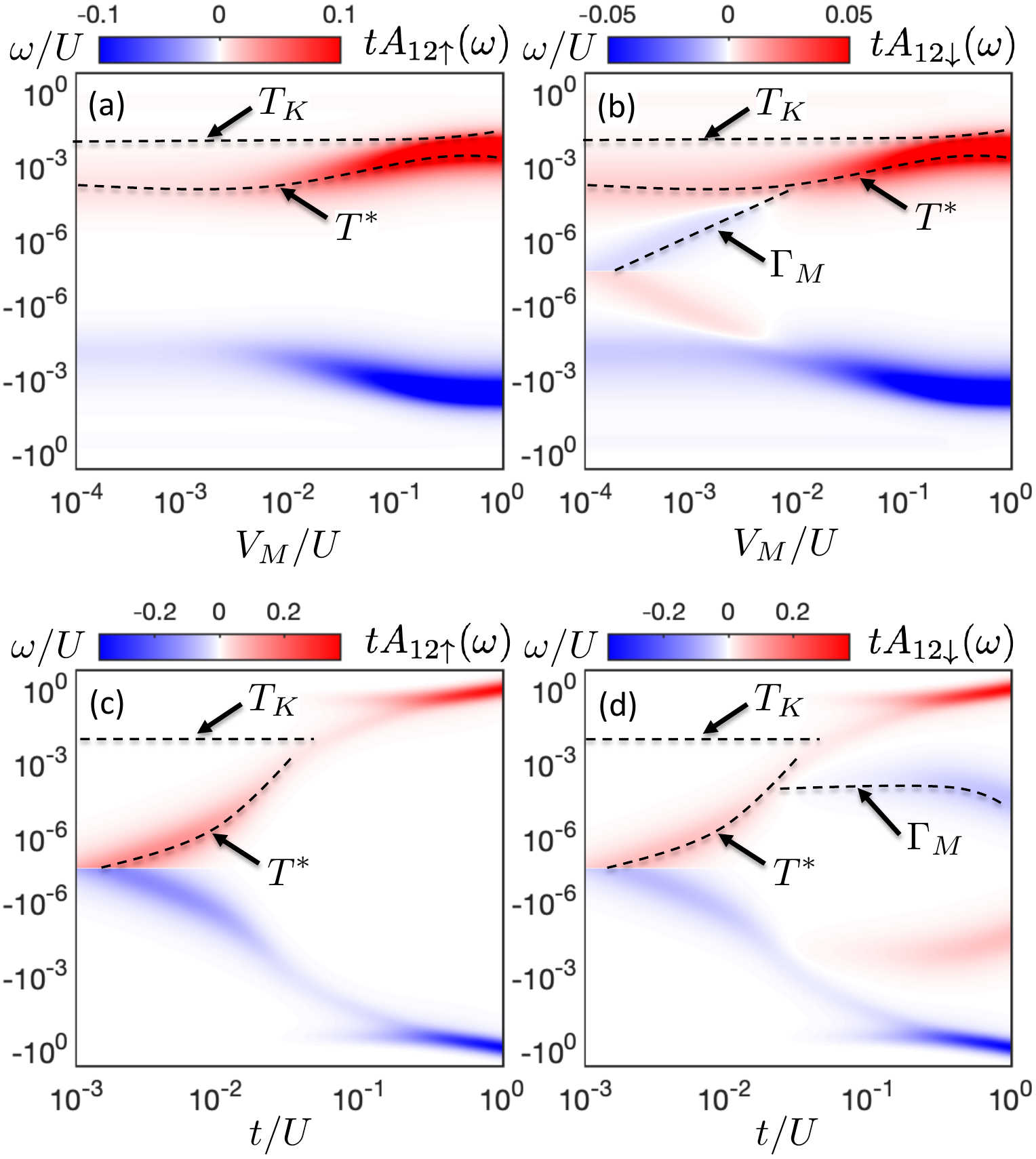}
	\caption{\label{Fig:A12}
		The spectral function $A_{12\sigma}(\omega)$
		for (a,c) spin-up and (b,d) spin-down components
		plotted  as function of  energy and (a,b) the coupling to Majorana wire $V_M$ for $t/U = 0.025$
		and (c,d) the hopping between the dots $t$ for $V_M/U = 0.01$.
		The other parameters are the same as in \fig{Fig:2}.
		Note the logarithmic scale on both axes.
	}
\end{figure}

\begin{figure}[t!]
	\includegraphics[width=1\columnwidth]{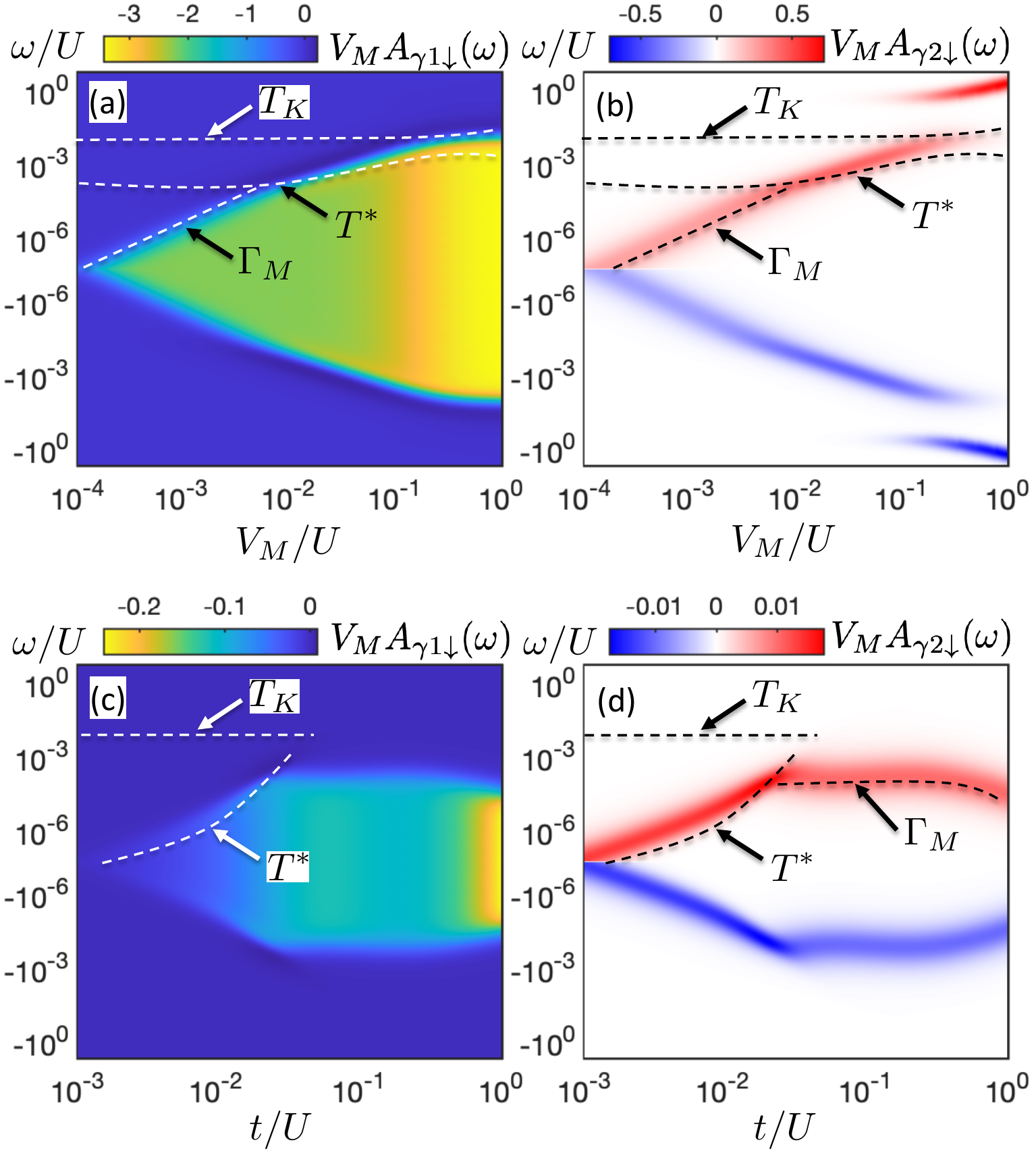}
	\caption{\label{Fig:Agamma}
		The spectral functions (a,c) $A_{\gamma 1\downarrow}(\omega)$
		and (b,d) $A_{\gamma 2\downarrow}(\omega)$
		plotted  as function of  energy and (a,b) the coupling to Majorana wire $V_M$ for $t/U = 0.025$
		and (c,d) the hopping between the dots $t$ for $V_M/U = 0.01$.
		The other parameters are the same as in \fig{Fig:2}.
		Note the logarithmic scale on both axes.
	}
\end{figure}

\begin{figure}[b!]
	\includegraphics[width=0.95\columnwidth]{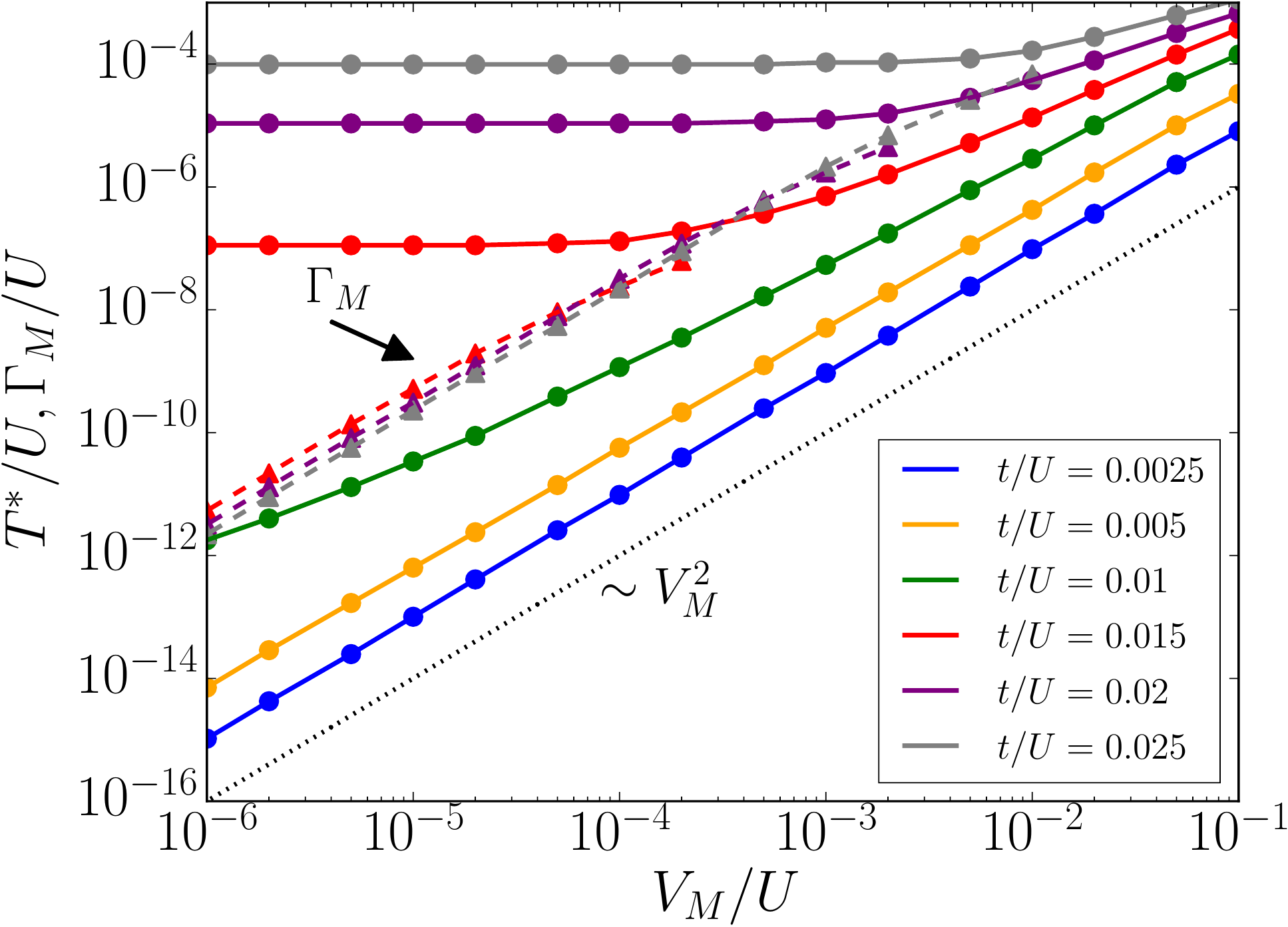}
	\caption{\label{Fig:GM}
		The second-stage Kondo temperature $T^*$ 
		and the Majorana energy scale $\Gamma_M$
		plotted as a function of $V_M$ for several values of the hopping
		between the quantum dots $t$.
		The other parameters are the same as in \fig{Fig:2}.
		The solid (dashed) line presents $T^*$ ($\Gamma_M$).
		Note the logarithmic scale in both axes.}
\end{figure}

Further insight into the leakage of Majorana states into the double dot setup
can be obtained from the analysis of off-diagonal spectral functions,
representing the correlations between the two quantum dots,
$A_{12\sigma}(\omega) = -\tfrac{1}{\pi} 
{\rm Im } \GFzub{d_{1\s}}{d_{2\s}^\dag}_\omega^R $,
as well as between the Majorana quasiparticle and 
electrons in the first and second quantum dot,
$A_{\gamma 1\downarrow}(\omega) = -\tfrac{1}{\pi} 
{\rm Im } \GFzub{\gamma}{d_{1\downarrow}^\dag}_\omega^R $
and
$A_{\gamma 2\downarrow}(\omega) = -\tfrac{1}{\pi} 
{\rm Im } \GFzub{\gamma}{d_{2\downarrow}^\dag}_\omega^R $,
respectively. These spectral functions
are presented in Figs.~\ref{Fig:A12} and \ref{Fig:Agamma}
for the same parameters as used in Figs. \ref{Fig:2}-\ref{Fig:3}.
The spectral function $A_{12\sigma}(\omega)$
has been normalized by the hopping $t$, whereas 
$A_{\gamma j\downarrow}(\omega)$ is normalized by $V_M$.
To facilitate the comparison, we have also included the corresponding dashed
lines presenting the relevant energy scales.

Let us begin the discussion with $A_{12\sigma}(\omega)$,
which describes the cross-correlations between the two quantum dots.
The spin-up component exhibits two patterns at energies approximately
corresponding to $\pm T^*$, which strongly depend on the value of hopping
and depart from the Fermi energy as $t$ grows, see 
Fig.~\ref{Fig:A12}(c). Finite coupling to topological wire
influences $A_{12\uparrow}(\omega)$ only for relatively
large $V_M\gtrsim t$, and shifts the resonances in $|A_{12\uparrow}(\omega)|$
to higher energies, see Fig.~\ref{Fig:A12}(a).
New features can be expected in the behavior of the 
opposite spin component, which is directly coupled to the Majorana mode.
Indeed, in Fig.~\ref{Fig:A12}(b) one can see that, as $V_M$ increases,
new resonances emerge at energy scale $\omega \approx \pm\Gamma_M$.
Moreover, similar features are visible in Fig.~\ref{Fig:A12}(d), which 
presents $A_{12\downarrow}(\omega)$ calculated while changing the hopping between the dots $t$.
It can be nicely seen that only if the hopping becomes
relatively large, Majorana mode can leak into the quantum dots.
We also note that the features related to the presence of Majorana 
modes result in the difference in the magnitude of 
spin components of  $A_{12\sigma}(\omega)$, such that 
$A_{12\uparrow}(\omega) / A_{12\downarrow}(\omega) = 2$.

To examine how exactly the Majorana state appears in the two dots,
in \fig{Fig:Agamma} we show the correlation function between the Majorana
quasiparticle and electrons in each of the quantum dots.
First of all, one can see that the behavior of 
$A_{\gamma 1\downarrow}(\omega)$
is completely different compared to that of $A_{\gamma 2\downarrow}(\omega)$,
although the features related to Majorana mode occur at approximately comparable energy scales in both spectral functions.
While $A_{\gamma 1\downarrow}(\omega)$ displays a plateau for $|\omega| \lesssim \Gamma_M$,
$A_{\gamma 2\downarrow}(\omega)$ exhibits a peak/dip for positive/negative energy
when $\omega \approx \pm \Gamma_M$. The position of these features grows
with $V_M$, see the first row of \fig{Fig:Agamma}. The corresponding spectral functions
plotted vs the hopping $t$ are shown in the second row of the figure.
It is seen that the energy associated with Majorana features 
grows with $t$, however, once $t\gtrsim V_M$, the dependence saturates,
see Figs. \ref{Fig:Agamma}(c) and (d).


Let us now examine how the magnitude of 
the second-stage Kondo temperature $T^*$ and the characteristic
Majorana energy scale $\Gamma_M$ depend on the coupling to topological wire.
There quantities as a function of $V_M$
are shown in \fig{Fig:GM} and are plotted for selected values of the hopping between the dots $t$.
$T^*$ was estimated as an energy scale at which
the spin-up spectral function drops to half of its maximum value
with decreasing the energy $\omega$.
On the other hand, the Majorana energy scale $\Gamma_M$
was determined from the energy at which the spin-down spectral function
drops from $1/(2\pi\Gamma)$ at $\omega\to 0$
to the half of its minimum value as the energy increases,
cf. Figs.~\ref{Fig:2}(a)-(b) and \ref{Fig:3}(a)-(b).
Note that in this way we can extract $\Gamma_M$
only for certain range of parameters, i.e. when $\Gamma_M\lesssim T^*$.

It can be nicely seen that $T^* \propto V_M^2$
for low values of the hopping between the dots, i.e. when the 
increase of $T^*$ is just due to the coupling to Majorana wire,
see e.g. the case of $t/U=0.0025$ in \fig{Fig:GM}.
However, when $t$ increases, a larger value of $V_M$
is needed in order to affect $T^*$. Nevertheless, once this happens,
$\Gamma_M$ again scales quadratically with $V_M$.
On the other hand, if $T^*$ is relatively large, increasing $V_M$
does not have any effect on the second-stage Kondo temperature,
see the curves for $t/U\geq 0.015$ for low values of $V_M$.
In other words, the influence of $V_M$ on the behavior
of $A_\uparrow(\omega)$ is negligible.
This is just contrary to $A_\downarrow(\omega)$,
which exhibits then new features due to
quantum interference with Majorana zero-energy mode,
resulting in an additional resonance at the Fermi energy.
As can be seen in \fig{Fig:GM} where $\Gamma_M$ is presented by dashed lines,
$\Gamma_M\propto V_M^2$, similarly to $T^*$.
Note also that the Majorana scale does not depend on
the hopping between the dots---the curves presenting
$\Gamma_M$ for different $t$ almost overlap, see \fig{Fig:GM}.

As follows from the discussion presented in this section,
the signatures of Majorana states are clearly visible
in both quantum dots.
Majorana correlations leak into the double dot giving rise to unique spectral features,
which in case of $A_{\sigma}(\omega)$ and $A_{2\sigma}(\omega)$
could be in principle probed with an STM tip.
Importantly, all the spin-down spectral functions reveal the energy
scales related to the Majorana zero-energy mode,
which allows to conclude that Majorana correlations are present in the whole nanostructure, not only at one of the quantum dots.
Nevertheless, because in our setup it is the spectral function of the first dot,
which is directly related to the conductance through
the system, cf. \eq{Eq:G}, from now on let us 
restrict ourselves to the discussion of the behavior of $A_\sigma(\omega)$.

\subsection{Linear conductance}

The interplay between the Majorana and Kondo physics
gives rise to well-resolved features visible
in the behavior of the linear conductance through the system.
First, let us discuss the temperature dependence of $G$,
whereas later on we turn to the analysis of the 
conductance dependence on the gate voltage.

\subsubsection{Temperature dependence}

\begin{figure}[t]
	\includegraphics[width=1\columnwidth]{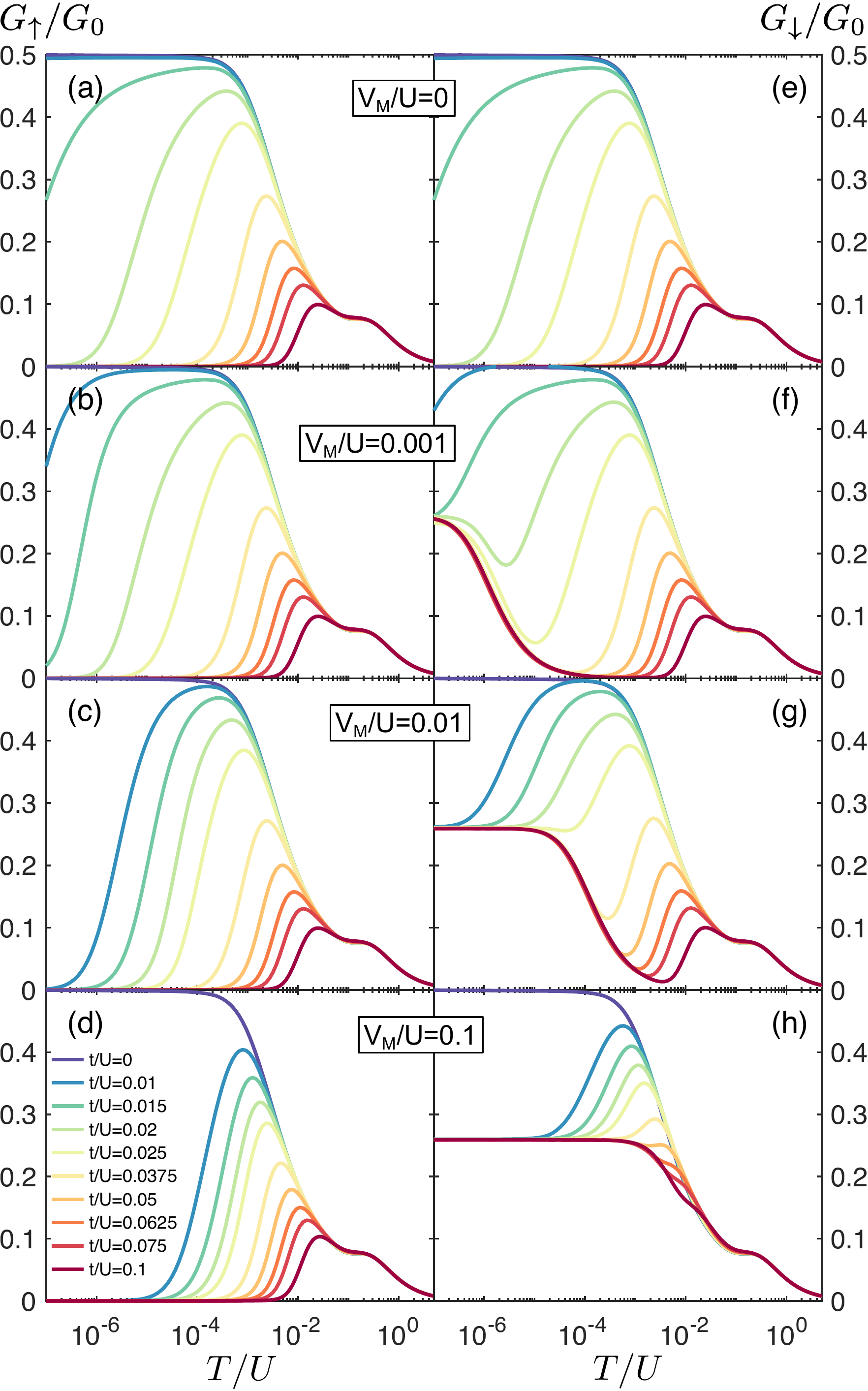}
	\caption{\label{Fig:4}
		The temperature dependence of the linear-response spin-resolved conductance
		through the double dot for (left column) spin-up and (right column) spin-down components
		calculated for different values of the hopping $t$ between the dots
		and the coupling to the Majorana wire $V_M$, as indicated.
		The other parameters are the same as in \fig{Fig:2}.}
\end{figure}

\begin{figure}[t!]
	\includegraphics[width=1\columnwidth]{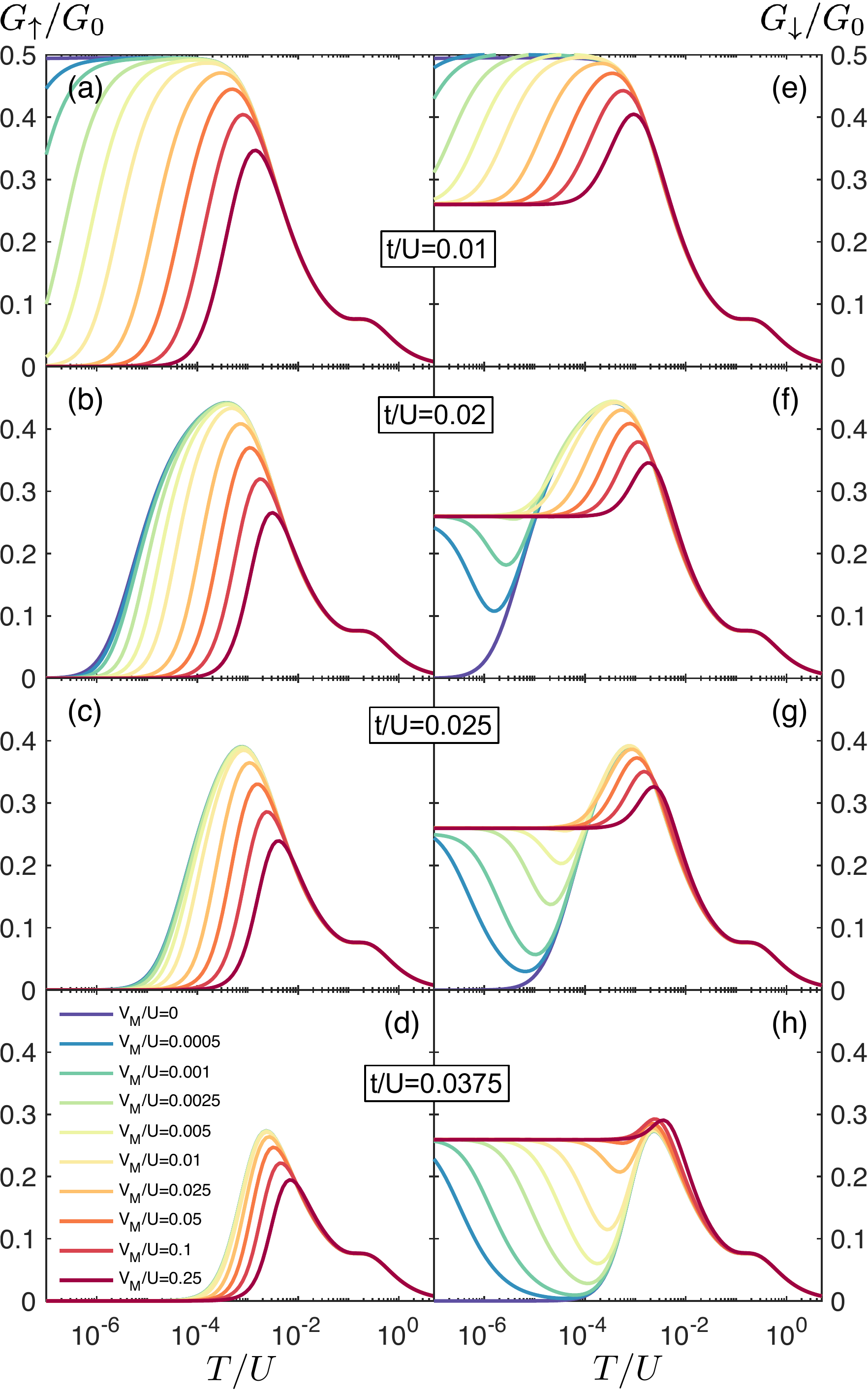}
	\caption{\label{Fig:5}
		The same as in \fig{Fig:4}, now plotted for selected values
		of the hopping between the dots $t$,
		while changing the coupling to the Majorana wire $V_M$.
	}
\end{figure}

The spin-resolved linear conductance as a function of temperature
calculated for different values of hopping between the dots
and the coupling to the Majorana wire is presented in Figs. 
\ref{Fig:4} and \ref{Fig:5}. While the first figure
displays $G(T)$ calculated for selected values of $V_M$
while tuning $t$, the second figure presents a complementary picture:
$G(T)$ determined for a few values of $t$ while changing $V_M$.
These figures nicely demonstrate the evolution
of the relevant energy scales in the system.

The spin-up component of the linear conductance
exhibits a typical nonmonotonic dependence due to the 
two-stage Kondo effect \cite{Cornaglia2005Feb}.
First, with lowering the temperature,
the conductance increases due to the Kondo effect,
however, around $T\approx T^*$, it starts to
drop due to the second stage of screening, at which 
the spin of the second dot becomes screened.
Since the second-stage Kondo temperature depends
strongly on the coupling to the Majorana wire, increasing $V_M$
results in an enhancement of $T^*$. As a consequence,
the maximum value of the conductance,
which develops for $T^*\lesssim T \lesssim T_K$
becomes reduced, see the left columns of 
Figs. \ref{Fig:4} and \ref{Fig:5}.

The enhancement of the second stage of Kondo screening
with raising $V_M$ (due to the increase of $T^*$)
is clearly visible in the left column of \fig{Fig:5}.
This enhancement  is more pronounced when the hopping
between the dots is relatively small. As shown in \fig{Fig:5}(a),
it is the coupling to Majorana wire that actually 
generates the second-stage of Kondo screening.
This is because $T^*$ for $t/U=10^{-2}$
is smaller than the energy scale presented in the figure.
However, when $t$ grows, larger values of coupling
to Majorana wire are needed in order to give rise to 
an increase of $T^*$. Nevertheless,
the advantageous impact of the coupling to the topological wire on $T^*$ is clearly visible.

On the other hand, the spin-down conductance
reveals much richer behavior due to the Majorana-Kondo interplay.
When the hopping between the dots is relatively small,
see \fig{Fig:5}(e) for $t/U=0.01$, $T^*$ for $V_M=0$
is smaller than the energy range considered in the figure
and a pronounced Kondo plateau is visible in the conductance.
Turning on the coupling to the Majorana wire,
results in a drop of  conductance to $G_\downarrow = (1/4)G_0$
at the characteristic energy scale $\w\approx \Gamma_M$, which grows with increasing $V_M$.
When the hopping between the dots becomes increased,
such that the second-stage Kondo screening 
can be visible in the behavior of the spin-up conductance,
one can observe an interplay between the 
Kondo effect and Majorana-induced quantum interference.
When $\Gamma_M\lesssim T^*$, a dip develops in the linear conductance
and $G_\down(T)$ exhibits two local maxima,
see e.g. Figs. \ref{Fig:4}(f)-(g)
and Figs. \ref{Fig:5}(f)-(g).
On the other hand, once $\Gamma_M\gtrsim T^*$,
the dip disappears and $G_\down(T)$
reaches $G_\downarrow = (1/4)G_0$ at very low temperatures.

\begin{figure}[t]
	\includegraphics[width=1\columnwidth]{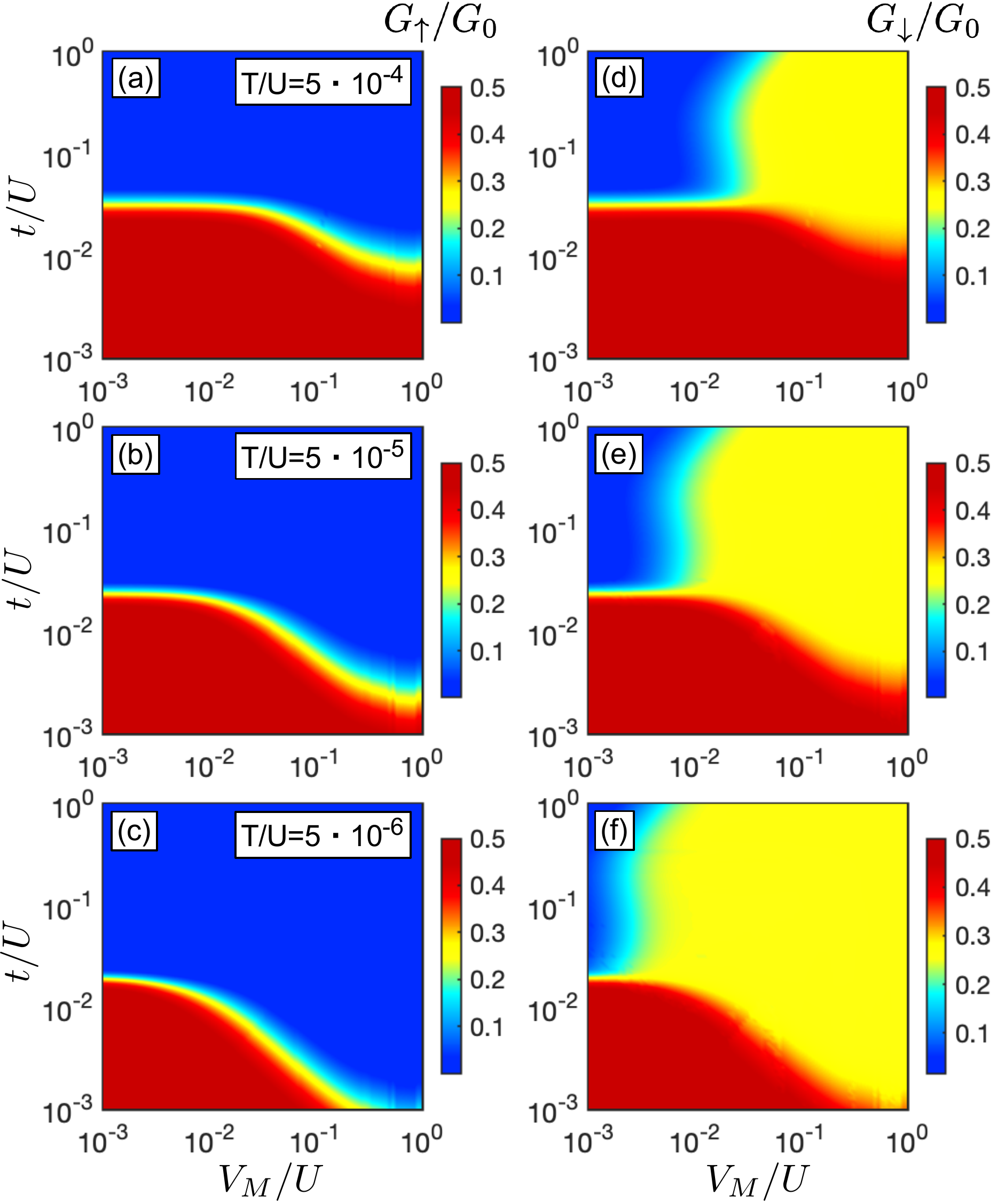}
	\caption{\label{Fig:6}
		The linear-response conductance calculated
		as a function of the hopping between the dots $t$
		and the coupling to the Majorana wire $V_M$
		for different temperatures, as indicated.
		The left (right) column presents
		the spin-up (spin-down) contribution.
		The other parameters are the same as in \fig{Fig:2}.
	Note the logarithmic scale for both $t$ and $V_M$.}
\end{figure}

While the dependence of conductance on temperature for $T\gtrsim T_K$
is almost the same for different values of $t$ and $V_M$
(for the range of parameters considered in Figs. \ref{Fig:4} and \ref{Fig:5}),
the behavior of low-temperature conductance is completely different.
Figure \ref{Fig:6} presents the 
linear-response conductance for both spin components
calculated at different temperatures
while tuning both $t$ and $V_M$.
Consider first the spin-up conductance for low-values of
$V_M$, see the left column of \fig{Fig:6}. By increasing the hopping between the dots,
the second-stage Kondo temperature becomes enhanced,
such that when $T^*\gtrsim T$, $G_\uparrow$
drops from $(1/2)G_0$ to $0$. Thus, the point
when this drop is observed is slightly different in each 
panel due to a different value of temperature $T$.
When the coupling to Majorana wire increases,
so does the second-stage Kondo temperature $T^*$,
such that the conductance drop is observed for smaller values of $t$.
In an extreme situation of very large $V_M$,
if the temperature is sufficiently low,
for all considered values of $t$ one has $T<T^*$,
such that the conductance stays suppressed due to the second-stage
Kondo effect, see \fig{Fig:6}(c) for $V_M\gtrsim U/10$.
A somewhat similar behavior can be observed
in the spin-down conductance component 
as far as the regions where $G_\uparrow = (1/2)G_0$ are concerned, see \fig{Fig:6}.
This is due to the fact that when the hopping between the two dots
is low, such that the second-stage Kondo effect does not develop,
the influence of the coupling to the Majorana wire is rather negligible
since the Majorana wire is coupled directly only to the second quantum dot.
It is therefore clear that the influence of presence of Majorana mode will be most revealed
in the parameter regime where the system exhibits the two-stage Kondo effect.
Consequently, one observes a completely different behavior
in the parameter space where $G_\uparrow\approx 0$,
cf. the left and right column of \fig{Fig:6}.
As can be clearly seen, with increasing $V_M$,
there is a value of $V_M$ at which the conductance 
increases from $0$ to $(1/4)G_0$. At lower temperatures,
smaller values of $V_M$ result in the corresponding change of conductance,
which is due to the fact that the condition $\Gamma_M \gtrsim T $
can be satisfied for smaller values of $V_M$.

\subsubsection{Gate voltage dependence}

\begin{figure}[t]
  \includegraphics[width=1\columnwidth]{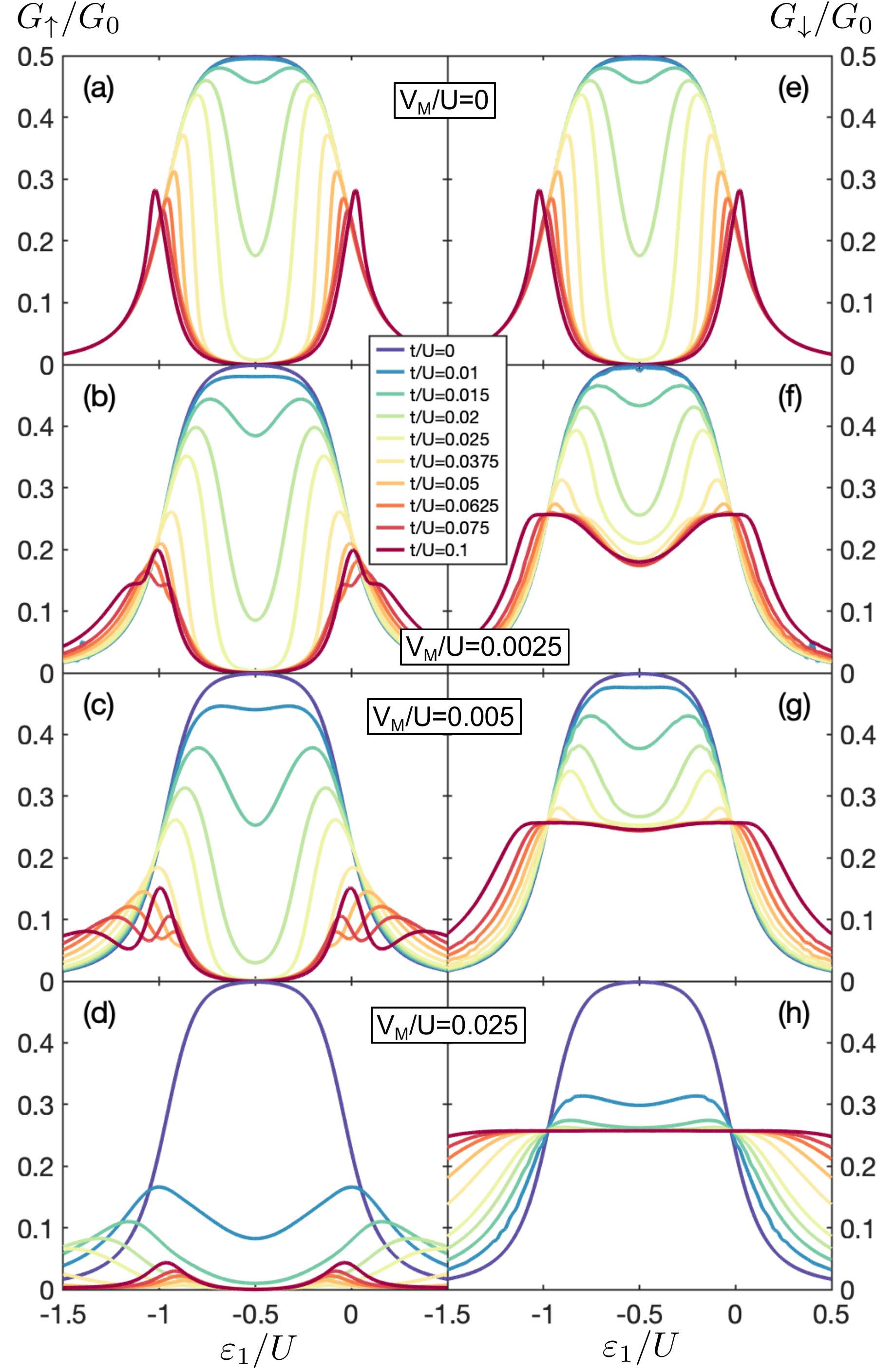}
  \caption{\label{Fig:7}
  The conductance plotted as a function of the
  position of the first quantum dot level $\e_1$
  for $\e_2 = -U/2$ and for selected values
  the coupling to Majorana mode $V_M$,
  while changing the hopping between the dots $t$.
  The other parameters are the same as in \fig{Fig:2}
  with $T=10^{-6} D$.
}
\end{figure}

\begin{figure}[t]
  \includegraphics[width=1\columnwidth]{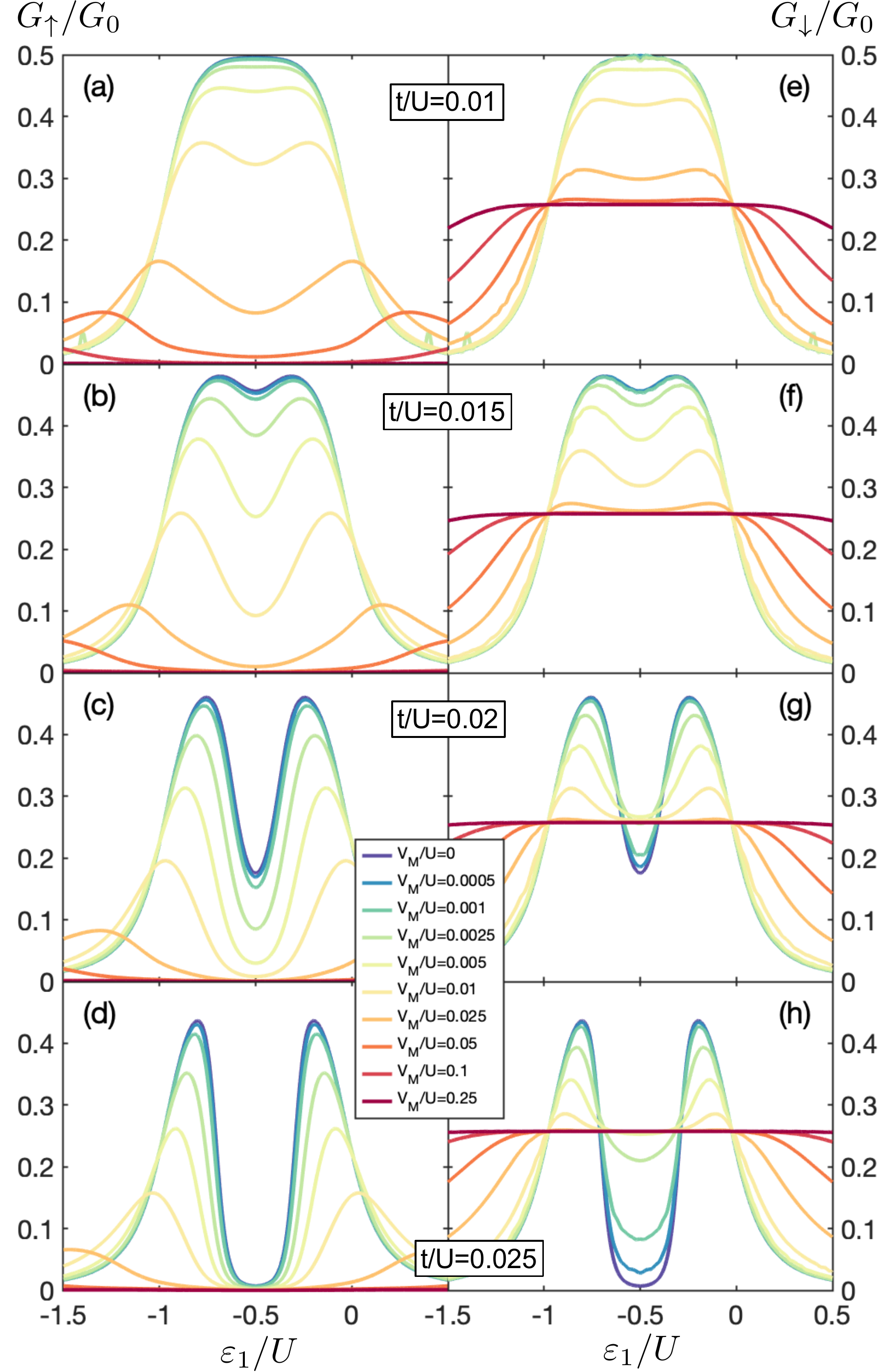}
  \caption{\label{Fig:8}
  The same as in \fig{Fig:6}, now plotted for selected values
  of hopping between the dots $t$,
  while changing the coupling to Majorana mode $V_M$.}
\end{figure}

Let us now discuss the gate voltage dependence
of the linear-response conductance. In the following
we consider the case when
the level of the first quantum dot is tuned,
while the level of the second dot is at half filling.
The spin-resolved conductance as a function
of $\e_1$ calculated for selected values of 
both $t$ and $V_M$ at extremely low yet non-zero temperature $T=10^{-6}U$ is shown in 
Figs. \ref{Fig:7} and \ref{Fig:8}.
Both spin-up and spin-down conductances
exhibit the Kondo plateau for small values of $t$ and $V_M$
in the transport regime where the first dot is singly occupied.
When the hopping between the dots increases, for $V_M=0$,
the Kondo plateau becomes distorted and 
the conductance suppression develops due to the two-stage Kondo effect,
see \fig{Fig:7}(a).
This suppression becomes more effective when
the coupling to Majorana wire is turned on, however,
then a clear difference between the spin components shows up.
In the spin-up channel, increasing $t$ and/or $V_M$,
generally results in larger suppression of the conductance
in the singly-occupied first dot regime, i.e. for $-U \lesssim \e_1 \lesssim 0$.
This is related to the corresponding increase 
of $T^*$, as already discussed in previous sections.
Similarly to the case $\e_1=\e_2=-U/2$ considered so far,
also for other values of $\e_1$ finite coupling
to the Majorana wire enhances $T^*$ and can
suppress the conductance through the system; see e.g.\fig{Fig:8}(a)

On the other hand, in the case of spin-down conductance 
one can see that once the second-stage Kondo effect comes into play
in the spin-up channel, i.e. suppression of conductance takes place,
$G_\downarrow$ reaches a fractional value of $(1/4)G_0$.
This is a direct fingerprint of the leakage of Majorana quasiparticle into the 
double-dot structure.
Consequently, with increasing $V_M$, the total conductance saturates 
at $(1/4)G_0$. Moreover, when the coupling to
the Majorana wire grows further, this value becomes stabilized in the whole region
of the gate voltage and the conductance hardly depends on the occupation
of the first quantum dot.
Similar effect has been predicted for single dots coupled 
to Majorana wire \cite{Lee2013Jun,Ruiz-Tijerina2015Mar,Wojcik2017,Gorski2018Oct}.
Here, we demonstrate that the Majorana zero-energy mode can leak through
the dot directly coupled to topological superconductor 
further into the nanostructure, and give
rise to fractional values of conductance.

\subsection{Short Majorana wire case}

\begin{figure}[t!]
	\includegraphics[width=1\columnwidth]{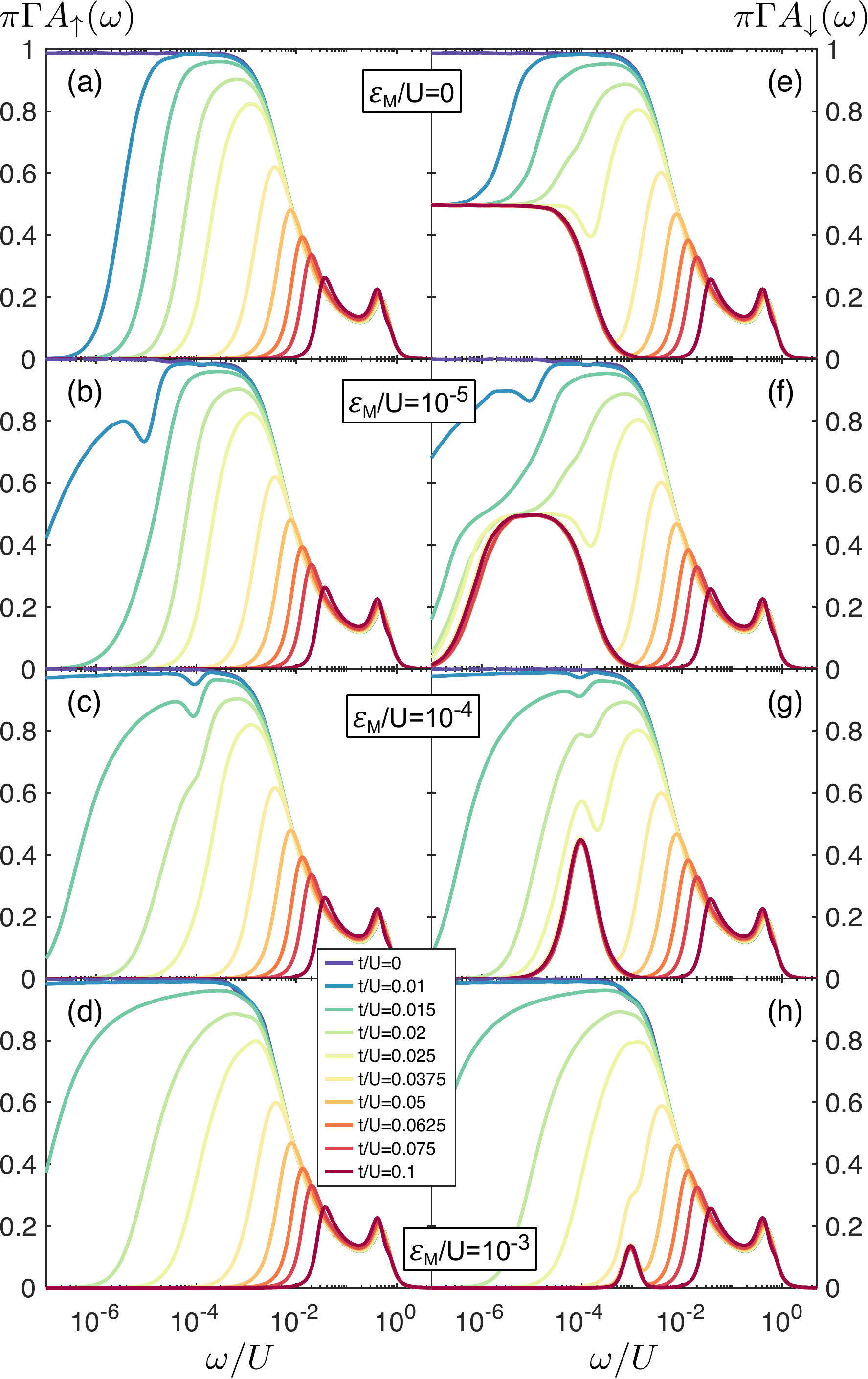}
	\caption{\label{Fig:9}
		The normalized spectral function $\pi\Gamma A_\sigma(\omega)$
		for the spin-up (left column) and spin-down (right column) components
		calculated for different values of the overlap between
		the Majorana zero-energy modes $\e_M$ 
		and the hopping $t$ between the two dots, as indicated.
		The parameters are the same as in \fig{Fig:2} with $V_M/U=0.01$.}
\end{figure}

\begin{figure}[t!]
	\includegraphics[width=1\columnwidth]{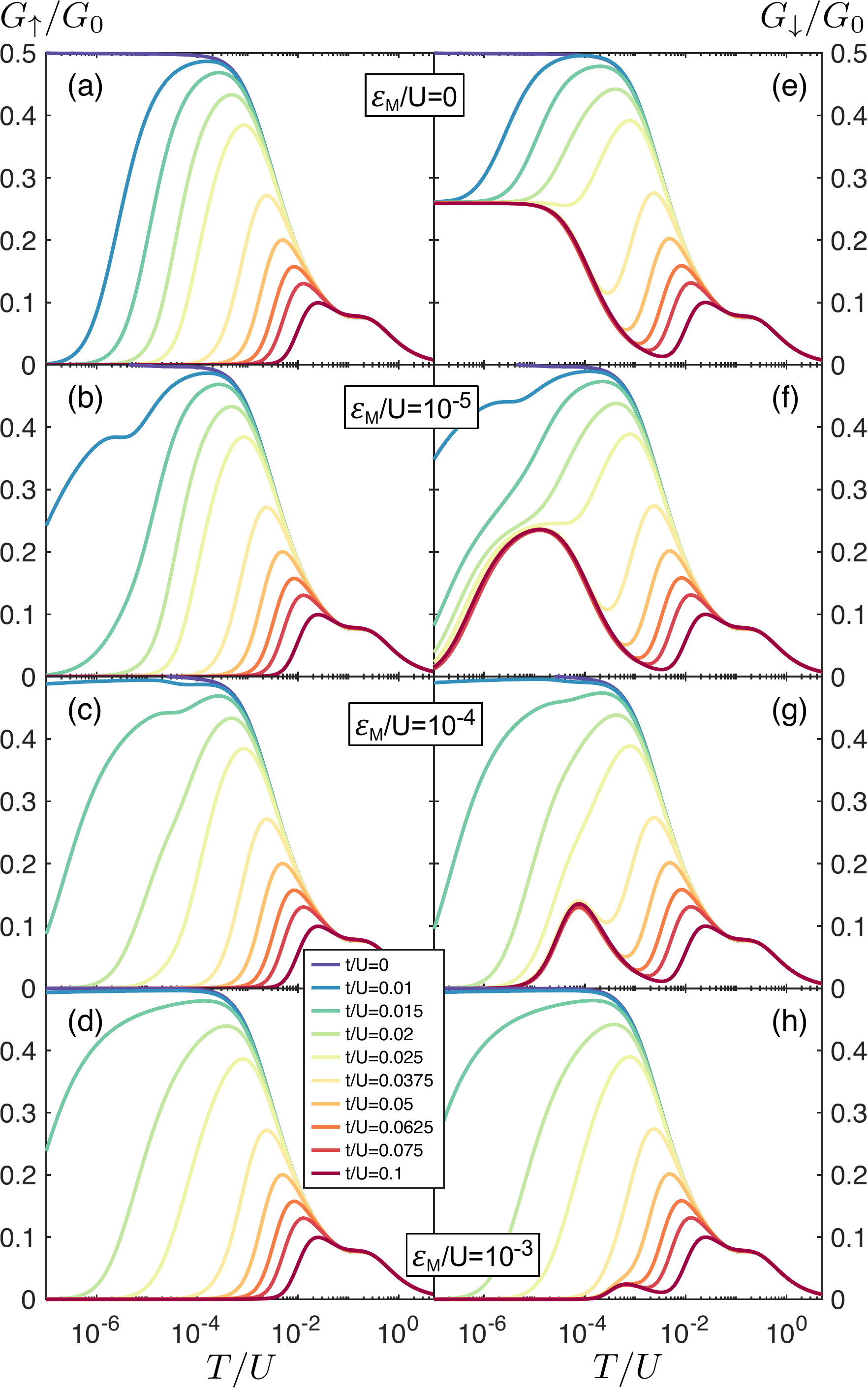}
	\caption{\label{Fig:10}
		The linear-response conductance calculated as a function
		of temperature for different values of the overlap
		$\e_M$ and the hopping between the dots $t$, as indicated.
		The left (right) column presents the spin-up (spin-down) component.
		The parameters are the same as in \fig{Fig:2} with $V_M/U=0.01$.}
\end{figure}

The interplay between the Majorana and Kondo 
correlations described in the previous sections can be greatly affected in the case of relatively short
topological superconducting wires. Then, a finite overlap between the 
wave functions of the two Majorana quasiparticles $\gamma_1$ and $\gamma_2$,
described by $\e_M$, can emerge.  
As shown in the following, such overlap has a strong influence on the 
quantum interference responsible for fractional values of the conductance.
In fact, such a strong dependence has already been
reported theoretically in the case of single quantum dots coupled
to external contacts and to the Majorana wire
\cite{Lee2013Jun,Lopez2014May,Weymann2017Jan,Weymann2017Apr}.
To examine the influence of the overlap $\e_M$
on the transport behavior of the considered double-dot-Majorana setup,
in \fig{Fig:9} we show the energy dependence of the spin-resolved
spectral function, while \fig{Fig:10} presents
the temperature dependence of the linear conductance through the system.
These figures were plotted based on calculations performed
for selected values of $\e_M$ while changing the hopping between the dots 
and for fixed coupling to Majorana wire, $V_M/U=0.01$.
The two figures are complementary in the sense
that the temperature dependence of the conductance
basically resembles the behavior of the spectral function,
except for the fact that some features are smeared out by thermal fluctuations.
The case of $\e_M=0$ presented in the first row of Figs. \ref{Fig:9} and \ref{Fig:10}
is just for reference, to help identifying the impact of finite overlap
on the behavior of $A_\sigma(\w)$ and $G_\sigma$.

The influence of $\e_M$ on the spin-up component can be 
observed in the left column of Figs. \ref{Fig:9} and \ref{Fig:10}.
A small local minimum can be seen at the energy scale 
corresponding to $\e_M$. Interestingly, below this energy scale
the interference with the Majorana mode becomes 
suppressed and the behavior of both $A_\uparrow(\w)$
and $G_\uparrow$ starts resembling that in the case of $V_M=0$.
This is especially visible for $\e_M/U=10^{-3}$,
which is presented in Figs. \ref{Fig:9}(d) and \ref{Fig:10}(d),
where the dependence of the quantities of interest 
is very similar to that depicted in Figs. \ref{Fig:2}(a) and \ref{Fig:4}(a), respectively.
Note also that this apparent switching off of the Majorana leakage
may lead to additional nonmonotonic behavior of the relevant spectral function
if the energy scales happen to fulfill $T^*(V_M=0) < \e_M < T^*(V_M)$;
see {\it e.g.} the curve for $t=0.01U$ in \fig{Fig:10}(b).

The destructive influence of the overlap $\e_M$ on the quantum interference
with Majorana mode is also visible in the spin-down component of both the spectral function
and the conductance, which are presented in the right columns of 
Figs. \ref{Fig:9} and \ref{Fig:10}. Now, a local minimum at the energy 
scale of $\e_M$ can also be observed, see e.g. the case for $\e_M/U=10^{-5}$.
Moreover, below this energy scale the behavior of $A_\downarrow(\w)$
and $G_\downarrow$ becomes comparable to that in the case of $V_M=0$,
i.e. the conductance becomes fully suppressed for large $t$
and $T\lesssim \e_M$. If the overlap is increased further, see the case of $\e_M/U=10^{-3}$,
the behavior of transport quantities resembles that in
the absence of coupling to the Majorana wire, except for 
an additional local maximum visible in both $A_\downarrow(\w)$
and $G_\downarrow$ at the energy scale corresponding to $\e_M$.
This result suggest that even strongly overlapping Majorana mode may partially leak into
the attached quantum dot.

\subsection{Majorana wire coupled to both spins}
\label{sec:2spins}

\begin{figure}[t]
	\includegraphics[width=1\columnwidth]{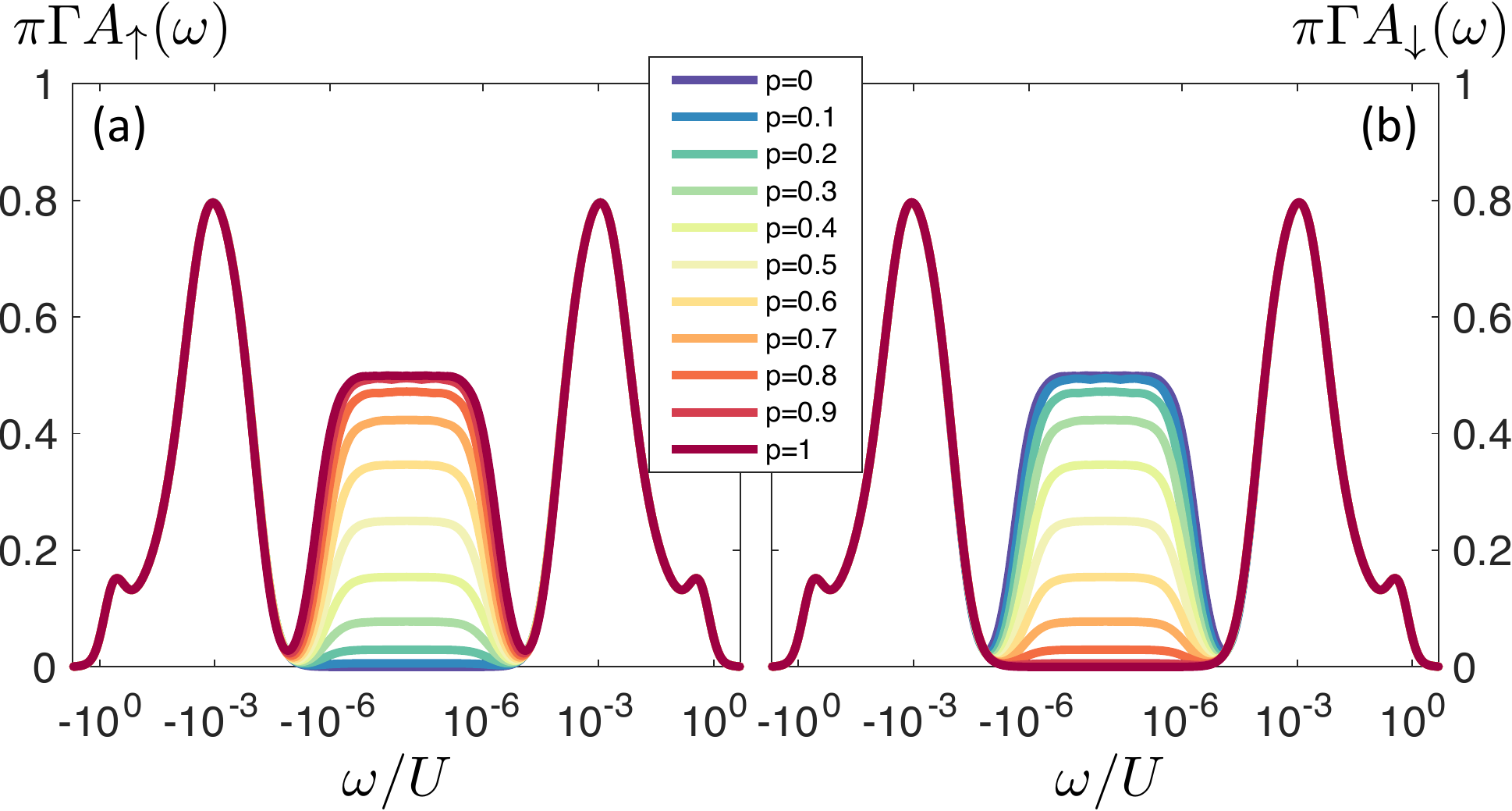}
	\caption{\label{Fig:12}
		The normalized spectral function $\pi\Gamma A_\sigma(\omega)$
		for (a) the spin-up and (b) spin-down components
		calculated for different values of the spin polarization $p$
		of the Majorana modes, as indicated.
		The parameters are the same as in \fig{Fig:2} with $V_M/U=0.001$
	   and $t/U=0.025$.}
\end{figure}

Finally, in this section we analyze the case when the Majorana quasiparticle
couples to both spin projections of the double dot. We thus assume that there
is a finite spin polarization of Majorana modes $p$ and express 
the double dot-Majorana coupling, cf. Eq.~(\ref{Eq:VM}), as \cite{Gorski2018Oct},
$\sum_\sigma V_{M\sigma}(d^\dagger_{2\sigma} - d_{2\sigma})(f^\dagger+f)$,
where $V_{M\downarrow} = (1-p) V_M$ and $V_{M\uparrow} = p V_M$.

The spin-resolved spectral functions calculated for different
values of polarization parameter $p$ are shown in \fig{Fig:12}.
This figure clearly illustrates the five-peak structure
in the behavior of the spectral function. The case of $p=0$
corresponds to no spin polarization of Majorana
quasiparticles---the Majorana wire couples only to spin-down electrons.
On the other hand, the case of $p=1$ corresponds to the situation
when Majorana modes are coupled only to spin-up electrons in the double dot.
One can see that now the spin-up spectral functions exhibits
all the features discussed previously for $A_\downarrow(\w)$.
When the spin polarization increases from $p=0$ to $p=1$,
the peak at the Fermi energy due to quantum interference with Majorana mode
is continuously transferred between the two spin components,
such that for $0<p<1$, the five-peak structure is visible in both spin-resolved
spectral functions. Note, however, that the total spectral function,
and consequently the total conductance, does not depend on $p$.

\section{Summary}
\label{sec:summary}

In this paper we have examined the transport behavior of a double quantum dot
in a T-shaped geometry, side-attached to a topological superconducting wire
hosting Majorana zero-energy modes. 
The considerations were performed by using the numerical renormalization group method,
which allowed us to accurately determine the behavior of the spin-resolved spectral functions
and the linear-response conductance of the system.
We have focused on the transport regime where the system exhibits
the two-stage Kondo effect and investigated the influence of the coupling 
to Majorana wire on this Kondo phenomenon, considering both long and short Majorana wire cases.
In the former case, the quantum interference with the Majorana
quasiparticle gives rise to a half-suppression of the second-stage of Kondo effect,
which results in a fractional value of the low-temperature conductance $G=(1/4)G_0$, where $G_0 = 2e^2/h$.
This phenomenon develops at a new energy scale $\Gamma_M \propto V_M^2$
associated with the coupling to Majorana wire described by hopping amplitude $V_M$.
We have shown that the Majorana-Kondo interplay can give rise
to an additional resonance in the local density of states for
energies lower than the Majorana scale $\Gamma_M$.
This can be interpreted as a Majorana mode leaking further
into the nanostructure, for it determines the low-temperature spectral properties
of the first quantum dot, while the Majorana wire is coupled to the second quantum dot.
At the same time, spectral density of the second dot is reduced by a half,
proving that both dots can exhibit signatures of the Majorana physics at the same time.
On the other hand, when there is an energy splitting $\e_M$ of Majorana modes due to a finite overlap 
of their wave functions, the quantum interference becomes suppressed
and the system exhibits the usual two-stage Kondo effect
for energies smaller than $\e_M$. Interestingly, 
both the conductance and spectral function
exhibit a local maximum at the energy scale corresponding to $\e_M$.

Our findings demonstrate that the low-temperature transport behavior
of T-shaped double quantum dots attached to Majorana wires
exhibits some unique features due to the leakage of Majorana quasiparticles
into the double dot system.
First of all, fractional values of the conductance develop in the Majorana-Kondo regime.
Moreover, the presence of topological superconductor increases the second-stage Kondo temperature through 
subtle renormalization effects, even though the relevant local exchange interaction is reduced.
On the other hand, in the case of relatively short wires, a local maximum develops in the conductance
for temperatures corresponding to the overlap between the two Majorana zero-energy modes.
These signatures provide further examples of unique transport behavior due to the presence 
of Majorana zero-energy modes.

We would also like to notice
that in the case of considered double quantum dot system
the presence of Majorana quasiparticles results in a huge relative change
of the low-temperature linear conductance [$G$ increases from $0$ to $(1/4)G_0$],
opposite to single dots where the relative change is much smaller
[$G$ drops from $G_0$ to $(3/4)G_0$].
The fractional conductance through T-shaped double quantum dot may thus
serve as another important fingerprint of the presence of Majorana
zero-energy modes in the system.

Finally, a comment on the intricate interplay of the three relevant energy scales
determining the transport behavior, i.e. the coupling to topological wire $V_M$,
the effective exchange interaction $\Jeff$ between the dots and the second-stage Kondo
temperature $T^*$, is due. Despite the fact that $V_M$ slightly decreases the bare 
exchange interaction between the quantum dots, accurate numerical analysis
has revealed an enhancement of the second-stage Kondo temperature.
This tendency may seem natural if one accepts the fact that finite $V_M$
facilitates the development of the Kondo effect in the single quantum dot case
\cite{Lee2013Jun,Ruiz-Tijerina2015Mar,Wojcik2017},
however, it could not be inferred from perturbative analysis.
It is thus not possible to a priori determine the fate of $T^*$,
in particular to know if this effect is sufficient to overcompensate
for the reduction of $\Jeff$ without numerical analysis.
Actually, only after careful and extensive examination of the system's
parameter space have we concluded that the tendency for increase
of $T^*$ with $V_M$ is indeed always the case.
However, the precise explanation why this is so needs further examination.

\begin{acknowledgments}
We thank J. Kroha, E. Vernek and T. Doma\'nski for stimulating discussions.
This work was supported by the National Science Centre
in Poland through the Project No. 2018/29/B/ST3/00937.
KPW acknowledges support from the National Science Centre in Poland 
through project No. 2015/19/N/ST3/01030
and from the Alexander von Humboldt Foundation.
PM acknowledges hospitality at University of Bonn.
Computing time at the Pozna\'n Supercomputing and Networking Center is appreciated.
\end{acknowledgments}



\begin{thebibliography}{66}%
	\makeatletter
	\providecommand \@ifxundefined [1]{%
		\@ifx{#1\undefined}
	}%
	\providecommand \@ifnum [1]{%
		\ifnum #1\expandafter \@firstoftwo
		\else \expandafter \@secondoftwo
		\fi
	}%
	\providecommand \@ifx [1]{%
		\ifx #1\expandafter \@firstoftwo
		\else \expandafter \@secondoftwo
		\fi
	}%
	\providecommand \natexlab [1]{#1}%
	\providecommand \enquote  [1]{``#1''}%
	\providecommand \bibnamefont  [1]{#1}%
	\providecommand \bibfnamefont [1]{#1}%
	\providecommand \citenamefont [1]{#1}%
	\providecommand \href@noop [0]{\@secondoftwo}%
	\providecommand \href [0]{\begingroup \@sanitize@url \@href}%
	\providecommand \@href[1]{\@@startlink{#1}\@@href}%
	\providecommand \@@href[1]{\endgroup#1\@@endlink}%
	\providecommand \@sanitize@url [0]{\catcode `\\12\catcode `\$12\catcode
		`\&12\catcode `\#12\catcode `\^12\catcode `\_12\catcode `\%12\relax}%
	\providecommand \@@startlink[1]{}%
	\providecommand \@@endlink[0]{}%
	\providecommand \url  [0]{\begingroup\@sanitize@url \@url }%
	\providecommand \@url [1]{\endgroup\@href {#1}{\urlprefix }}%
	\providecommand \urlprefix  [0]{URL }%
	\providecommand \Eprint [0]{\href }%
	\providecommand \doibase [0]{https://doi.org/}%
	\providecommand \selectlanguage [0]{\@gobble}%
	\providecommand \bibinfo  [0]{\@secondoftwo}%
	\providecommand \bibfield  [0]{\@secondoftwo}%
	\providecommand \translation [1]{[#1]}%
	\providecommand \BibitemOpen [0]{}%
	\providecommand \bibitemStop [0]{}%
	\providecommand \bibitemNoStop [0]{.\EOS\space}%
	\providecommand \EOS [0]{\spacefactor3000\relax}%
	\providecommand \BibitemShut  [1]{\csname bibitem#1\endcsname}%
	\let\auto@bib@innerbib\@empty
	\bibitem [{\citenamefont {Hasan}\ and\ \citenamefont
		{Kane}(2010)}]{Hasan2010Nov}%
	\BibitemOpen
	\bibfield  {author} {\bibinfo {author} {\bibfnamefont {M.~Z.}\ \bibnamefont
			{Hasan}}\ and\ \bibinfo {author} {\bibfnamefont {C.~L.}\ \bibnamefont
			{Kane}},\ }\bibfield  {title} {\bibinfo {title} {{Colloquium: Topological
				insulators}},\ }\href {https://doi.org/10.1103/RevModPhys.82.3045} {\bibfield
		{journal} {\bibinfo  {journal} {Rev. Mod. Phys.}\ }\textbf {\bibinfo
			{volume} {82}},\ \bibinfo {pages} {3045} (\bibinfo {year}
		{2010})}\BibitemShut {NoStop}%
	\bibitem [{\citenamefont {Qi}\ and\ \citenamefont {Zhang}(2011)}]{Qi2011Oct}%
	\BibitemOpen
	\bibfield  {author} {\bibinfo {author} {\bibfnamefont {X.-L.}\ \bibnamefont
			{Qi}}\ and\ \bibinfo {author} {\bibfnamefont {S.-C.}\ \bibnamefont {Zhang}},\
	}\bibfield  {title} {\bibinfo {title} {{Topological insulators and
				superconductors}},\ }\href {https://doi.org/10.1103/RevModPhys.83.1057}
	{\bibfield  {journal} {\bibinfo  {journal} {Rev. Mod. Phys.}\ }\textbf
		{\bibinfo {volume} {83}},\ \bibinfo {pages} {1057} (\bibinfo {year}
		{2011})}\BibitemShut {NoStop}%
	\bibitem [{\citenamefont {Wang}\ and\ \citenamefont
		{Zhang}(2017)}]{Wang2017Oct}%
	\BibitemOpen
	\bibfield  {author} {\bibinfo {author} {\bibfnamefont {J.}~\bibnamefont
			{Wang}}\ and\ \bibinfo {author} {\bibfnamefont {S.-C.}\ \bibnamefont
			{Zhang}},\ }\bibfield  {title} {\bibinfo {title} {{Topological states of
				condensed matter}},\ }\href {https://doi.org/10.1038/nmat5012} {\bibfield
		{journal} {\bibinfo  {journal} {Nat. Mater.}\ }\textbf {\bibinfo {volume}
			{16}},\ \bibinfo {pages} {1062} (\bibinfo {year} {2017})}\BibitemShut
	{NoStop}%
	\bibitem [{\citenamefont {Nayak}\ \emph {et~al.}(2008)\citenamefont {Nayak},
		\citenamefont {Simon}, \citenamefont {Stern}, \citenamefont {Freedman},\ and\
		\citenamefont {Das~Sarma}}]{Nayak2008Sep}%
	\BibitemOpen
	\bibfield  {author} {\bibinfo {author} {\bibfnamefont {C.}~\bibnamefont
			{Nayak}}, \bibinfo {author} {\bibfnamefont {S.~H.}\ \bibnamefont {Simon}},
		\bibinfo {author} {\bibfnamefont {A.}~\bibnamefont {Stern}}, \bibinfo
		{author} {\bibfnamefont {M.}~\bibnamefont {Freedman}},\ and\ \bibinfo
		{author} {\bibfnamefont {S.}~\bibnamefont {Das~Sarma}},\ }\bibfield  {title}
	{\bibinfo {title} {{Non-Abelian anyons and topological quantum
				computation}},\ }\href {https://doi.org/10.1103/RevModPhys.80.1083}
	{\bibfield  {journal} {\bibinfo  {journal} {Rev. Mod. Phys.}\ }\textbf
		{\bibinfo {volume} {80}},\ \bibinfo {pages} {1083} (\bibinfo {year}
		{2008})}\BibitemShut {NoStop}%
	\bibitem [{\citenamefont {Majorana}(1937)}]{Majorana1937Apr}%
	\BibitemOpen
	\bibfield  {author} {\bibinfo {author} {\bibfnamefont {E.}~\bibnamefont
			{Majorana}},\ }\bibfield  {title} {\bibinfo {title} {{Teoria simmetrica
				dell{'}elettrone e del positrone}},\ }\href
	{https://doi.org/10.1007/BF02961314} {\bibfield  {journal} {\bibinfo
			{journal} {Nuovo Cim.}\ }\textbf {\bibinfo {volume} {14}},\ \bibinfo {pages}
		{171} (\bibinfo {year} {1937})}\BibitemShut {NoStop}%
	\bibitem [{\citenamefont {Kitaev}(2003)}]{Kitaev2003Jan}%
	\BibitemOpen
	\bibfield  {author} {\bibinfo {author} {\bibfnamefont {A.~{\relax Yu}.}\
			\bibnamefont {Kitaev}},\ }\bibfield  {title} {\bibinfo {title}
		{{Fault-tolerant quantum computation by anyons}},\ }\href
	{https://doi.org/10.1016/S0003-4916(02)00018-0} {\bibfield  {journal}
		{\bibinfo  {journal} {Ann. Phys.}\ }\textbf {\bibinfo {volume} {303}},\
		\bibinfo {pages} {2} (\bibinfo {year} {2003})}\BibitemShut {NoStop}%
	\bibitem [{\citenamefont {Alicea}(2012)}]{Alicea2012Jun}%
	\BibitemOpen
	\bibfield  {author} {\bibinfo {author} {\bibfnamefont {J.}~\bibnamefont
			{Alicea}},\ }\bibfield  {title} {\bibinfo {title} {{New directions in the
				pursuit of Majorana fermions in solid state systems}},\ }\href
	{https://doi.org/10.1088/0034-4885/75/7/076501} {\bibfield  {journal}
		{\bibinfo  {journal} {Rep. Prog. Phys.}\ }\textbf {\bibinfo {volume} {75}},\
		\bibinfo {pages} {076501} (\bibinfo {year} {2012})}\BibitemShut {NoStop}%
	\bibitem [{\citenamefont {Mourik}\ \emph {et~al.}(2012)\citenamefont {Mourik},
		\citenamefont {Zuo}, \citenamefont {Frolov}, \citenamefont {Plissard},
		\citenamefont {Bakkers},\ and\ \citenamefont {Kouwenhoven}}]{Mourik2012May}%
	\BibitemOpen
	\bibfield  {author} {\bibinfo {author} {\bibfnamefont {V.}~\bibnamefont
			{Mourik}}, \bibinfo {author} {\bibfnamefont {K.}~\bibnamefont {Zuo}},
		\bibinfo {author} {\bibfnamefont {S.~M.}\ \bibnamefont {Frolov}}, \bibinfo
		{author} {\bibfnamefont {S.~R.}\ \bibnamefont {Plissard}}, \bibinfo {author}
		{\bibfnamefont {E.~P. A.~M.}\ \bibnamefont {Bakkers}},\ and\ \bibinfo
		{author} {\bibfnamefont {L.~P.}\ \bibnamefont {Kouwenhoven}},\ }\bibfield
	{title} {\bibinfo {title} {{Signatures of Majorana Fermions in Hybrid
				Superconductor-Semiconductor Nanowire Devices}},\ }\href
	{https://doi.org/10.1126/science.1222360} {\bibfield  {journal} {\bibinfo
			{journal} {Science}\ }\textbf {\bibinfo {volume} {336}},\ \bibinfo {pages}
		{1003} (\bibinfo {year} {2012})}\BibitemShut {NoStop}%
	\bibitem [{\citenamefont {Deng}\ \emph {et~al.}(2012)\citenamefont {Deng},
		\citenamefont {Yu}, \citenamefont {Huang}, \citenamefont {Larsson},
		\citenamefont {Caroff},\ and\ \citenamefont {Xu}}]{Deng2012Nov}%
	\BibitemOpen
	\bibfield  {author} {\bibinfo {author} {\bibfnamefont {M.~T.}\ \bibnamefont
			{Deng}}, \bibinfo {author} {\bibfnamefont {C.~L.}\ \bibnamefont {Yu}},
		\bibinfo {author} {\bibfnamefont {G.~Y.}\ \bibnamefont {Huang}}, \bibinfo
		{author} {\bibfnamefont {M.}~\bibnamefont {Larsson}}, \bibinfo {author}
		{\bibfnamefont {P.}~\bibnamefont {Caroff}},\ and\ \bibinfo {author}
		{\bibfnamefont {H.~Q.}\ \bibnamefont {Xu}},\ }\bibfield  {title} {\bibinfo
		{title} {{Anomalous Zero-Bias Conductance Peak in a Nb{\textendash}InSb
				Nanowire{\textendash}Nb Hybrid Device}},\ }\bibfield  {journal} {\bibinfo
		{journal} {American Chemical Society}\ }\href
	{https://doi.org/10.1021/nl303758w} {10.1021/nl303758w} (\bibinfo {year}
	{2012})\BibitemShut {NoStop}%
	\bibitem [{\citenamefont {Das}\ \emph {et~al.}(2012)\citenamefont {Das},
		\citenamefont {Ronen}, \citenamefont {Most}, \citenamefont {Oreg},
		\citenamefont {Heiblum},\ and\ \citenamefont {Shtrikman}}]{Das2012Nov}%
	\BibitemOpen
	\bibfield  {author} {\bibinfo {author} {\bibfnamefont {A.}~\bibnamefont
			{Das}}, \bibinfo {author} {\bibfnamefont {Y.}~\bibnamefont {Ronen}}, \bibinfo
		{author} {\bibfnamefont {Y.}~\bibnamefont {Most}}, \bibinfo {author}
		{\bibfnamefont {Y.}~\bibnamefont {Oreg}}, \bibinfo {author} {\bibfnamefont
			{M.}~\bibnamefont {Heiblum}},\ and\ \bibinfo {author} {\bibfnamefont
			{H.}~\bibnamefont {Shtrikman}},\ }\bibfield  {title} {\bibinfo {title}
		{{Zero-bias peaks and splitting in an Al{\textendash}InAs nanowire
				topological superconductor as a signature of Majorana fermions}},\ }\href
	{https://doi.org/10.1038/nphys2479} {\bibfield  {journal} {\bibinfo
			{journal} {Nat. Phys.}\ }\textbf {\bibinfo {volume} {8}},\ \bibinfo {pages}
		{887} (\bibinfo {year} {2012})}\BibitemShut {NoStop}%
	\bibitem [{\citenamefont {Albrecht}\ \emph {et~al.}(2016)\citenamefont
		{Albrecht}, \citenamefont {Higginbotham}, \citenamefont {Madsen},
		\citenamefont {Kuemmeth}, \citenamefont {Jespersen}, \citenamefont
		{Nyg{\aa}rd}, \citenamefont {Krogstrup},\ and\ \citenamefont
		{Marcus}}]{Albrecht2016Mar}%
	\BibitemOpen
	\bibfield  {author} {\bibinfo {author} {\bibfnamefont {S.~M.}\ \bibnamefont
			{Albrecht}}, \bibinfo {author} {\bibfnamefont {A.~P.}\ \bibnamefont
			{Higginbotham}}, \bibinfo {author} {\bibfnamefont {M.}~\bibnamefont
			{Madsen}}, \bibinfo {author} {\bibfnamefont {F.}~\bibnamefont {Kuemmeth}},
		\bibinfo {author} {\bibfnamefont {T.~S.}\ \bibnamefont {Jespersen}}, \bibinfo
		{author} {\bibfnamefont {J.}~\bibnamefont {Nyg{\aa}rd}}, \bibinfo {author}
		{\bibfnamefont {P.}~\bibnamefont {Krogstrup}},\ and\ \bibinfo {author}
		{\bibfnamefont {C.~M.}\ \bibnamefont {Marcus}},\ }\bibfield  {title}
	{\bibinfo {title} {{Exponential protection of zero modes in Majorana
				islands}},\ }\href {https://doi.org/10.1038/nature17162} {\bibfield
		{journal} {\bibinfo  {journal} {Nature}\ }\textbf {\bibinfo {volume} {531}},\
		\bibinfo {pages} {206} (\bibinfo {year} {2016})}\BibitemShut {NoStop}%
	\bibitem [{\citenamefont {Deng}\ \emph {et~al.}(2016)\citenamefont {Deng},
		\citenamefont {Vaitiek{\ifmmode\dot{e}\else\.{e}\fi}nas}, \citenamefont
		{Hansen}, \citenamefont {Danon}, \citenamefont {Leijnse}, \citenamefont
		{Flensberg}, \citenamefont {Nyg{\aa}rd}, \citenamefont {Krogstrup},\ and\
		\citenamefont {Marcus}}]{Deng2016Dec}%
	\BibitemOpen
	\bibfield  {author} {\bibinfo {author} {\bibfnamefont {M.~T.}\ \bibnamefont
			{Deng}}, \bibinfo {author} {\bibfnamefont {S.}~\bibnamefont
			{Vaitiek{\ifmmode\dot{e}\else\.{e}\fi}nas}}, \bibinfo {author} {\bibfnamefont
			{E.~B.}\ \bibnamefont {Hansen}}, \bibinfo {author} {\bibfnamefont
			{J.}~\bibnamefont {Danon}}, \bibinfo {author} {\bibfnamefont
			{M.}~\bibnamefont {Leijnse}}, \bibinfo {author} {\bibfnamefont
			{K.}~\bibnamefont {Flensberg}}, \bibinfo {author} {\bibfnamefont
			{J.}~\bibnamefont {Nyg{\aa}rd}}, \bibinfo {author} {\bibfnamefont
			{P.}~\bibnamefont {Krogstrup}},\ and\ \bibinfo {author} {\bibfnamefont
			{C.~M.}\ \bibnamefont {Marcus}},\ }\bibfield  {title} {\bibinfo {title}
		{{Majorana bound state in a coupled quantum-dot hybrid-nanowire system}},\
	}\href {https://doi.org/10.1126/science.aaf3961} {\bibfield  {journal}
		{\bibinfo  {journal} {Science}\ }\textbf {\bibinfo {volume} {354}},\ \bibinfo
		{pages} {1557} (\bibinfo {year} {2016})}\BibitemShut {NoStop}%
	\bibitem [{\citenamefont {Deng}\ \emph {et~al.}(2018)\citenamefont {Deng},
		\citenamefont {Vaitiek{\ifmmode\dot{e}\else\.{e}\fi}nas}, \citenamefont
		{Prada}, \citenamefont {San-Jose}, \citenamefont {Nyg{\aa}rd}, \citenamefont
		{Krogstrup}, \citenamefont {Aguado},\ and\ \citenamefont
		{Marcus}}]{Deng2018Aug}%
	\BibitemOpen
	\bibfield  {author} {\bibinfo {author} {\bibfnamefont {M.-T.}\ \bibnamefont
			{Deng}}, \bibinfo {author} {\bibfnamefont {S.}~\bibnamefont
			{Vaitiek{\ifmmode\dot{e}\else\.{e}\fi}nas}}, \bibinfo {author} {\bibfnamefont
			{E.}~\bibnamefont {Prada}}, \bibinfo {author} {\bibfnamefont
			{P.}~\bibnamefont {San-Jose}}, \bibinfo {author} {\bibfnamefont
			{J.}~\bibnamefont {Nyg{\aa}rd}}, \bibinfo {author} {\bibfnamefont
			{P.}~\bibnamefont {Krogstrup}}, \bibinfo {author} {\bibfnamefont
			{R.}~\bibnamefont {Aguado}},\ and\ \bibinfo {author} {\bibfnamefont {C.~M.}\
			\bibnamefont {Marcus}},\ }\bibfield  {title} {\bibinfo {title} {{Nonlocality
				of Majorana modes in hybrid nanowires}},\ }\href
	{https://doi.org/10.1103/PhysRevB.98.085125} {\bibfield  {journal} {\bibinfo
			{journal} {Phys. Rev. B}\ }\textbf {\bibinfo {volume} {98}},\ \bibinfo
		{pages} {085125} (\bibinfo {year} {2018})}\BibitemShut {NoStop}%
	\bibitem [{\citenamefont {Zhang}\ \emph {et~al.}(2018)\citenamefont {Zhang},
		\citenamefont {Liu}, \citenamefont {Gazibegovic}, \citenamefont {Xu},
		\citenamefont {Logan}, \citenamefont {Wang}, \citenamefont {van Loo},
		\citenamefont {Bommer}, \citenamefont {de~Moor}, \citenamefont {Car},
		\citenamefont {Op~het Veld}, \citenamefont {van Veldhoven}, \citenamefont
		{Koelling}, \citenamefont {Verheijen}, \citenamefont {Pendharkar},
		\citenamefont {Pennachio}, \citenamefont {Shojaei}, \citenamefont {Lee},
		\citenamefont {Palmstr{\o}m}, \citenamefont {Bakkers}, \citenamefont
		{Sarma},\ and\ \citenamefont {Kouwenhoven}}]{Zhang2018Mar}%
	\BibitemOpen
	\bibfield  {author} {\bibinfo {author} {\bibfnamefont {H.}~\bibnamefont
			{Zhang}}, \bibinfo {author} {\bibfnamefont {C.-X.}\ \bibnamefont {Liu}},
		\bibinfo {author} {\bibfnamefont {S.}~\bibnamefont {Gazibegovic}}, \bibinfo
		{author} {\bibfnamefont {D.}~\bibnamefont {Xu}}, \bibinfo {author}
		{\bibfnamefont {J.~A.}\ \bibnamefont {Logan}}, \bibinfo {author}
		{\bibfnamefont {G.}~\bibnamefont {Wang}}, \bibinfo {author} {\bibfnamefont
			{N.}~\bibnamefont {van Loo}}, \bibinfo {author} {\bibfnamefont {J.~D.~S.}\
			\bibnamefont {Bommer}}, \bibinfo {author} {\bibfnamefont {M.~W.~A.}\
			\bibnamefont {de~Moor}}, \bibinfo {author} {\bibfnamefont {D.}~\bibnamefont
			{Car}}, \bibinfo {author} {\bibfnamefont {R.~L.~M.}\ \bibnamefont {Op~het
				Veld}}, \bibinfo {author} {\bibfnamefont {P.~J.}\ \bibnamefont {van
				Veldhoven}}, \bibinfo {author} {\bibfnamefont {S.}~\bibnamefont {Koelling}},
		\bibinfo {author} {\bibfnamefont {M.~A.}\ \bibnamefont {Verheijen}}, \bibinfo
		{author} {\bibfnamefont {M.}~\bibnamefont {Pendharkar}}, \bibinfo {author}
		{\bibfnamefont {D.~J.}\ \bibnamefont {Pennachio}}, \bibinfo {author}
		{\bibfnamefont {B.}~\bibnamefont {Shojaei}}, \bibinfo {author} {\bibfnamefont
			{J.~S.}\ \bibnamefont {Lee}}, \bibinfo {author} {\bibfnamefont {C.~J.}\
			\bibnamefont {Palmstr{\o}m}}, \bibinfo {author} {\bibfnamefont {E.~P. A.~M.}\
			\bibnamefont {Bakkers}}, \bibinfo {author} {\bibfnamefont {S.~D.}\
			\bibnamefont {Sarma}},\ and\ \bibinfo {author} {\bibfnamefont {L.~P.}\
			\bibnamefont {Kouwenhoven}},\ }\bibfield  {title} {\bibinfo {title}
		{{Quantized Majorana conductance}},\ }\href
	{https://doi.org/10.1038/nature26142} {\bibfield  {journal} {\bibinfo
			{journal} {Nature}\ }\textbf {\bibinfo {volume} {556}},\ \bibinfo {pages}
		{74} (\bibinfo {year} {2018})}\BibitemShut {NoStop}%
	\bibitem [{\citenamefont {Lutchyn}\ \emph {et~al.}(2018)\citenamefont
		{Lutchyn}, \citenamefont {Bakkers}, \citenamefont {Kouwenhoven},
		\citenamefont {Krogstrup}, \citenamefont {Marcus},\ and\ \citenamefont
		{Oreg}}]{Lutchyn2018May}%
	\BibitemOpen
	\bibfield  {author} {\bibinfo {author} {\bibfnamefont {R.~M.}\ \bibnamefont
			{Lutchyn}}, \bibinfo {author} {\bibfnamefont {E.~P. A.~M.}\ \bibnamefont
			{Bakkers}}, \bibinfo {author} {\bibfnamefont {L.~P.}\ \bibnamefont
			{Kouwenhoven}}, \bibinfo {author} {\bibfnamefont {P.}~\bibnamefont
			{Krogstrup}}, \bibinfo {author} {\bibfnamefont {C.~M.}\ \bibnamefont
			{Marcus}},\ and\ \bibinfo {author} {\bibfnamefont {Y.}~\bibnamefont {Oreg}},\
	}\bibfield  {title} {\bibinfo {title} {{Majorana zero modes in
				superconductor{\textendash}semiconductor heterostructures}},\ }\href
	{https://doi.org/10.1038/s41578-018-0003-1} {\bibfield  {journal} {\bibinfo
			{journal} {Nat. Rev. Mater.}\ }\textbf {\bibinfo {volume} {3}},\ \bibinfo
		{pages} {52} (\bibinfo {year} {2018})}\BibitemShut {NoStop}%
	\bibitem [{\citenamefont {G{\ifmmode\ddot{u}\else\"{u}\fi}l}\ \emph
		{et~al.}(2018)\citenamefont {G{\ifmmode\ddot{u}\else\"{u}\fi}l},
		\citenamefont {Zhang}, \citenamefont {Bommer}, \citenamefont {de~Moor},
		\citenamefont {Car}, \citenamefont {Plissard}, \citenamefont {Bakkers},
		\citenamefont {Geresdi}, \citenamefont {Watanabe}, \citenamefont
		{Taniguchi},\ and\ \citenamefont {Kouwenhoven}}]{Gul2018Jan}%
	\BibitemOpen
	\bibfield  {author} {\bibinfo {author} {\bibfnamefont
			{{\ifmmode\ddot{O}\else\"{O}\fi}.}~\bibnamefont
			{G{\ifmmode\ddot{u}\else\"{u}\fi}l}}, \bibinfo {author} {\bibfnamefont
			{H.}~\bibnamefont {Zhang}}, \bibinfo {author} {\bibfnamefont {J.~D.~S.}\
			\bibnamefont {Bommer}}, \bibinfo {author} {\bibfnamefont {M.~W.~A.}\
			\bibnamefont {de~Moor}}, \bibinfo {author} {\bibfnamefont {D.}~\bibnamefont
			{Car}}, \bibinfo {author} {\bibfnamefont {S.~R.}\ \bibnamefont {Plissard}},
		\bibinfo {author} {\bibfnamefont {E.~P. A.~M.}\ \bibnamefont {Bakkers}},
		\bibinfo {author} {\bibfnamefont {A.}~\bibnamefont {Geresdi}}, \bibinfo
		{author} {\bibfnamefont {K.}~\bibnamefont {Watanabe}}, \bibinfo {author}
		{\bibfnamefont {T.}~\bibnamefont {Taniguchi}},\ and\ \bibinfo {author}
		{\bibfnamefont {L.~P.}\ \bibnamefont {Kouwenhoven}},\ }\bibfield  {title}
	{\bibinfo {title} {{Ballistic Majorana nanowire devices}},\ }\href
	{https://doi.org/10.1038/s41565-017-0032-8} {\bibfield  {journal} {\bibinfo
			{journal} {Nat. Nanotechnol.}\ }\textbf {\bibinfo {volume} {13}},\ \bibinfo
		{pages} {192} (\bibinfo {year} {2018})}\BibitemShut {NoStop}%
	\bibitem [{\citenamefont {Liu}\ and\ \citenamefont
		{Baranger}(2011)}]{Liu2011Nov}%
	\BibitemOpen
	\bibfield  {author} {\bibinfo {author} {\bibfnamefont {D.~E.}\ \bibnamefont
			{Liu}}\ and\ \bibinfo {author} {\bibfnamefont {H.~U.}\ \bibnamefont
			{Baranger}},\ }\bibfield  {title} {\bibinfo {title} {{Detecting a
				Majorana-fermion zero mode using a quantum dot}},\ }\href
	{https://doi.org/10.1103/PhysRevB.84.201308} {\bibfield  {journal} {\bibinfo
			{journal} {Phys. Rev. B}\ }\textbf {\bibinfo {volume} {84}},\ \bibinfo
		{pages} {201308} (\bibinfo {year} {2011})}\BibitemShut {NoStop}%
	\bibitem [{\citenamefont {Leijnse}\ and\ \citenamefont
		{Flensberg}(2011)}]{Leijnse2011Oct}%
	\BibitemOpen
	\bibfield  {author} {\bibinfo {author} {\bibfnamefont {M.}~\bibnamefont
			{Leijnse}}\ and\ \bibinfo {author} {\bibfnamefont {K.}~\bibnamefont
			{Flensberg}},\ }\bibfield  {title} {\bibinfo {title} {{Scheme to measure
				Majorana fermion lifetimes using a quantum dot}},\ }\href
	{https://doi.org/10.1103/PhysRevB.84.140501} {\bibfield  {journal} {\bibinfo
			{journal} {Phys. Rev. B}\ }\textbf {\bibinfo {volume} {84}},\ \bibinfo
		{pages} {140501} (\bibinfo {year} {2011})}\BibitemShut {NoStop}%
	\bibitem [{\citenamefont {Cao}\ \emph {et~al.}(2012)\citenamefont {Cao},
		\citenamefont {Wang}, \citenamefont {Xiong}, \citenamefont {Gong},\ and\
		\citenamefont {Li}}]{Cao2012Sep}%
	\BibitemOpen
	\bibfield  {author} {\bibinfo {author} {\bibfnamefont {Y.}~\bibnamefont
			{Cao}}, \bibinfo {author} {\bibfnamefont {P.}~\bibnamefont {Wang}}, \bibinfo
		{author} {\bibfnamefont {G.}~\bibnamefont {Xiong}}, \bibinfo {author}
		{\bibfnamefont {M.}~\bibnamefont {Gong}},\ and\ \bibinfo {author}
		{\bibfnamefont {X.-Q.}\ \bibnamefont {Li}},\ }\bibfield  {title} {\bibinfo
		{title} {{Probing the existence and dynamics of Majorana fermion via
				transport through a quantum dot}},\ }\href
	{https://doi.org/10.1103/PhysRevB.86.115311} {\bibfield  {journal} {\bibinfo
			{journal} {Phys. Rev. B}\ }\textbf {\bibinfo {volume} {86}},\ \bibinfo
		{pages} {115311} (\bibinfo {year} {2012})}\BibitemShut {NoStop}%
	\bibitem [{\citenamefont {Gong}\ \emph {et~al.}(2014)\citenamefont {Gong},
		\citenamefont {Zhang}, \citenamefont {Li}, \citenamefont {Yi},\ and\
		\citenamefont {Zheng}}]{Gong2014Jun}%
	\BibitemOpen
	\bibfield  {author} {\bibinfo {author} {\bibfnamefont {W.-J.}\ \bibnamefont
			{Gong}}, \bibinfo {author} {\bibfnamefont {S.-F.}\ \bibnamefont {Zhang}},
		\bibinfo {author} {\bibfnamefont {Z.-C.}\ \bibnamefont {Li}}, \bibinfo
		{author} {\bibfnamefont {G.}~\bibnamefont {Yi}},\ and\ \bibinfo {author}
		{\bibfnamefont {Y.-S.}\ \bibnamefont {Zheng}},\ }\bibfield  {title} {\bibinfo
		{title} {{Detection of a Majorana fermion zero mode by a T-shaped quantum-dot
				structure}},\ }\href {https://doi.org/10.1103/PhysRevB.89.245413} {\bibfield
		{journal} {\bibinfo  {journal} {Phys. Rev. B}\ }\textbf {\bibinfo {volume}
			{89}},\ \bibinfo {pages} {245413} (\bibinfo {year} {2014})}\BibitemShut
	{NoStop}%
	\bibitem [{\citenamefont {Liu}\ \emph {et~al.}(2015)\citenamefont {Liu},
		\citenamefont {Cheng},\ and\ \citenamefont {Lutchyn}}]{Liu2015Feb}%
	\BibitemOpen
	\bibfield  {author} {\bibinfo {author} {\bibfnamefont {D.~E.}\ \bibnamefont
			{Liu}}, \bibinfo {author} {\bibfnamefont {M.}~\bibnamefont {Cheng}},\ and\
		\bibinfo {author} {\bibfnamefont {R.~M.}\ \bibnamefont {Lutchyn}},\
	}\bibfield  {title} {\bibinfo {title} {{Probing Majorana physics in
				quantum-dot shot-noise experiments}},\ }\href
	{https://doi.org/10.1103/PhysRevB.91.081405} {\bibfield  {journal} {\bibinfo
			{journal} {Phys. Rev. B}\ }\textbf {\bibinfo {volume} {91}},\ \bibinfo
		{pages} {081405} (\bibinfo {year} {2015})}\BibitemShut {NoStop}%
	\bibitem [{\citenamefont {Weymann}\ and\ \citenamefont
		{W{\ifmmode\acute{o}\else\'{o}\fi}jcik}(2017)}]{Weymann2017Apr}%
	\BibitemOpen
	\bibfield  {author} {\bibinfo {author} {\bibfnamefont {I.}~\bibnamefont
			{Weymann}}\ and\ \bibinfo {author} {\bibfnamefont {K.~P.}\ \bibnamefont
			{W{\ifmmode\acute{o}\else\'{o}\fi}jcik}},\ }\bibfield  {title} {\bibinfo
		{title} {{Transport properties of a hybrid Majorana wire-quantum dot system
				with ferromagnetic contacts}},\ }\href
	{https://doi.org/10.1103/PhysRevB.95.155427} {\bibfield  {journal} {\bibinfo
			{journal} {Phys. Rev. B}\ }\textbf {\bibinfo {volume} {95}},\ \bibinfo
		{pages} {155427} (\bibinfo {year} {2017})}\BibitemShut {NoStop}%
	\bibitem [{\citenamefont {Liu}\ \emph {et~al.}(2017)\citenamefont {Liu},
		\citenamefont {Sau}, \citenamefont {Stanescu},\ and\ \citenamefont
		{Das~Sarma}}]{Liu2017Aug}%
	\BibitemOpen
	\bibfield  {author} {\bibinfo {author} {\bibfnamefont {C.-X.}\ \bibnamefont
			{Liu}}, \bibinfo {author} {\bibfnamefont {J.~D.}\ \bibnamefont {Sau}},
		\bibinfo {author} {\bibfnamefont {T.~D.}\ \bibnamefont {Stanescu}},\ and\
		\bibinfo {author} {\bibfnamefont {S.}~\bibnamefont {Das~Sarma}},\ }\bibfield
	{title} {\bibinfo {title} {{Andreev bound states versus Majorana bound states
				in quantum dot-nanowire-superconductor hybrid structures: Trivial versus
				topological zero-bias conductance peaks}},\ }\href
	{https://doi.org/10.1103/PhysRevB.96.075161} {\bibfield  {journal} {\bibinfo
			{journal} {Phys. Rev. B}\ }\textbf {\bibinfo {volume} {96}},\ \bibinfo
		{pages} {075161} (\bibinfo {year} {2017})}\BibitemShut {NoStop}%
	\bibitem [{\citenamefont {Prada}\ \emph {et~al.}(2017)\citenamefont {Prada},
		\citenamefont {Aguado},\ and\ \citenamefont {San-Jose}}]{Prada2017Aug}%
	\BibitemOpen
	\bibfield  {author} {\bibinfo {author} {\bibfnamefont {E.}~\bibnamefont
			{Prada}}, \bibinfo {author} {\bibfnamefont {R.}~\bibnamefont {Aguado}},\ and\
		\bibinfo {author} {\bibfnamefont {P.}~\bibnamefont {San-Jose}},\ }\bibfield
	{title} {\bibinfo {title} {{Measuring Majorana nonlocality and spin structure
				with a quantum dot}},\ }\href {https://doi.org/10.1103/PhysRevB.96.085418}
	{\bibfield  {journal} {\bibinfo  {journal} {Phys. Rev. B}\ }\textbf {\bibinfo
			{volume} {96}},\ \bibinfo {pages} {085418} (\bibinfo {year}
		{2017})}\BibitemShut {NoStop}%
	\bibitem [{\citenamefont {Ptok}\ \emph {et~al.}(2017)\citenamefont {Ptok},
		\citenamefont {Kobia{\l}ka},\ and\ \citenamefont
		{Doma{\ifmmode\acute{n}\else\'{n}\fi}ski}}]{Ptok2017Nov}%
	\BibitemOpen
	\bibfield  {author} {\bibinfo {author} {\bibfnamefont {A.}~\bibnamefont
			{Ptok}}, \bibinfo {author} {\bibfnamefont {A.}~\bibnamefont {Kobia{\l}ka}},\
		and\ \bibinfo {author} {\bibfnamefont {T.}~\bibnamefont
			{Doma{\ifmmode\acute{n}\else\'{n}\fi}ski}},\ }\bibfield  {title} {\bibinfo
		{title} {{Controlling the bound states in a quantum-dot hybrid nanowire}},\
	}\href {https://doi.org/10.1103/PhysRevB.96.195430} {\bibfield  {journal}
		{\bibinfo  {journal} {Phys. Rev. B}\ }\textbf {\bibinfo {volume} {96}},\
		\bibinfo {pages} {195430} (\bibinfo {year} {2017})}\BibitemShut {NoStop}%
	\bibitem [{\citenamefont {G{\ifmmode\acute{o}\else\'{o}\fi}rski}\ \emph
		{et~al.}(2018)\citenamefont {G{\ifmmode\acute{o}\else\'{o}\fi}rski},
		\citenamefont {Bara{\ifmmode\acute{n}\else\'{n}\fi}ski}, \citenamefont
		{Weymann},\ and\ \citenamefont
		{Doma{\ifmmode\acute{n}\else\'{n}\fi}ski}}]{Gorski2018Oct}%
	\BibitemOpen
	\bibfield  {author} {\bibinfo {author} {\bibfnamefont {G.}~\bibnamefont
			{G{\ifmmode\acute{o}\else\'{o}\fi}rski}}, \bibinfo {author} {\bibfnamefont
			{J.}~\bibnamefont {Bara{\ifmmode\acute{n}\else\'{n}\fi}ski}}, \bibinfo
		{author} {\bibfnamefont {I.}~\bibnamefont {Weymann}},\ and\ \bibinfo {author}
		{\bibfnamefont {T.}~\bibnamefont {Doma{\ifmmode\acute{n}\else\'{n}\fi}ski}},\
	}\bibfield  {title} {\bibinfo {title} {{Interplay between correlations and
				Majorana mode in proximitized quantum dot}},\ }\href
	{https://doi.org/10.1038/s41598-018-33529-1} {\bibfield  {journal} {\bibinfo
			{journal} {Sci. Rep.}\ }\textbf {\bibinfo {volume} {8}},\ \bibinfo {pages}
		{1} (\bibinfo {year} {2018})}\BibitemShut {NoStop}%
	\bibitem [{\citenamefont {Stenger}\ \emph {et~al.}(2018)\citenamefont
		{Stenger}, \citenamefont {Woods}, \citenamefont {Frolov},\ and\ \citenamefont
		{Stanescu}}]{Stenger2018Aug}%
	\BibitemOpen
	\bibfield  {author} {\bibinfo {author} {\bibfnamefont {J.~P.~T.}\
			\bibnamefont {Stenger}}, \bibinfo {author} {\bibfnamefont {B.~D.}\
			\bibnamefont {Woods}}, \bibinfo {author} {\bibfnamefont {S.~M.}\ \bibnamefont
			{Frolov}},\ and\ \bibinfo {author} {\bibfnamefont {T.~D.}\ \bibnamefont
			{Stanescu}},\ }\bibfield  {title} {\bibinfo {title} {{Control and detection
				of Majorana bound states in quantum dot arrays}},\ }\href
	{https://doi.org/10.1103/PhysRevB.98.085407} {\bibfield  {journal} {\bibinfo
			{journal} {Phys. Rev. B}\ }\textbf {\bibinfo {volume} {98}},\ \bibinfo
		{pages} {085407} (\bibinfo {year} {2018})}\BibitemShut {NoStop}%
	\bibitem [{\citenamefont {Cifuentes}\ and\ \citenamefont
		{Da~Silva}(2019)}]{Cifuentes2019Aug}%
	\BibitemOpen
	\bibfield  {author} {\bibinfo {author} {\bibfnamefont {J.~D.}\ \bibnamefont
			{Cifuentes}}\ and\ \bibinfo {author} {\bibfnamefont {L.~G. G. V.~D.}\
			\bibnamefont {Da~Silva}},\ }\bibfield  {title} {\bibinfo {title}
		{{Manipulating Majorana zero modes in double quantum dots}},\ }\href
	{https://doi.org/10.1103/PhysRevB.100.085429} {\bibfield  {journal} {\bibinfo
			{journal} {Phys. Rev. B}\ }\textbf {\bibinfo {volume} {100}},\ \bibinfo
		{pages} {085429} (\bibinfo {year} {2019})}\BibitemShut {NoStop}%
	\bibitem [{\citenamefont {Silva}\ \emph {et~al.}(2020)\citenamefont {Silva},
		\citenamefont {Da~Silva},\ and\ \citenamefont {Vernek}}]{Vernek2019}%
	\BibitemOpen
	\bibfield  {author} {\bibinfo {author} {\bibfnamefont {J.~F.}\ \bibnamefont
			{Silva}}, \bibinfo {author} {\bibfnamefont {L.~G. G. V.~D.}\ \bibnamefont
			{Da~Silva}},\ and\ \bibinfo {author} {\bibfnamefont {E.}~\bibnamefont
			{Vernek}},\ }\bibfield  {title} {\bibinfo {title} {{Robustness of the Kondo
				effect in a quantum dot coupled to Majorana zero modes}},\ }\href
	{https://doi.org/10.1103/PhysRevB.101.075428} {\bibfield  {journal} {\bibinfo
			{journal} {Phys. Rev. B}\ }\textbf {\bibinfo {volume} {101}},\ \bibinfo
		{pages} {075428} (\bibinfo {year} {2020})}\BibitemShut {NoStop}%
	\bibitem [{\citenamefont {Vernek}\ \emph {et~al.}(2014)\citenamefont {Vernek},
		\citenamefont {Penteado}, \citenamefont {Seridonio},\ and\ \citenamefont
		{Egues}}]{Vernek2014Apr}%
	\BibitemOpen
	\bibfield  {author} {\bibinfo {author} {\bibfnamefont {E.}~\bibnamefont
			{Vernek}}, \bibinfo {author} {\bibfnamefont {P.~H.}\ \bibnamefont
			{Penteado}}, \bibinfo {author} {\bibfnamefont {A.~C.}\ \bibnamefont
			{Seridonio}},\ and\ \bibinfo {author} {\bibfnamefont {J.~C.}\ \bibnamefont
			{Egues}},\ }\bibfield  {title} {\bibinfo {title} {{Subtle leakage of a
				Majorana mode into a quantum dot}},\ }\href
	{https://doi.org/10.1103/PhysRevB.89.165314} {\bibfield  {journal} {\bibinfo
			{journal} {Phys. Rev. B}\ }\textbf {\bibinfo {volume} {89}},\ \bibinfo
		{pages} {165314} (\bibinfo {year} {2014})}\BibitemShut {NoStop}%
	\bibitem [{\citenamefont {Ruiz-Tijerina}\ \emph {et~al.}(2015)\citenamefont
		{Ruiz-Tijerina}, \citenamefont {Vernek}, \citenamefont {Dias~da Silva},\ and\
		\citenamefont {Egues}}]{Ruiz-Tijerina2015Mar}%
	\BibitemOpen
	\bibfield  {author} {\bibinfo {author} {\bibfnamefont {D.~A.}\ \bibnamefont
			{Ruiz-Tijerina}}, \bibinfo {author} {\bibfnamefont {E.}~\bibnamefont
			{Vernek}}, \bibinfo {author} {\bibfnamefont {L.~G. G.~V.}\ \bibnamefont
			{Dias~da Silva}},\ and\ \bibinfo {author} {\bibfnamefont {J.~C.}\
			\bibnamefont {Egues}},\ }\bibfield  {title} {\bibinfo {title} {{Interaction
				effects on a Majorana zero mode leaking into a quantum dot}},\ }\href
	{https://doi.org/10.1103/PhysRevB.91.115435} {\bibfield  {journal} {\bibinfo
			{journal} {Phys. Rev. B}\ }\textbf {\bibinfo {volume} {91}},\ \bibinfo
		{pages} {115435} (\bibinfo {year} {2015})}\BibitemShut {NoStop}%
	\bibitem [{\citenamefont {Golub}\ \emph {et~al.}(2011)\citenamefont {Golub},
		\citenamefont {Kuzmenko},\ and\ \citenamefont {Avishai}}]{Golub2011Oct}%
	\BibitemOpen
	\bibfield  {author} {\bibinfo {author} {\bibfnamefont {A.}~\bibnamefont
			{Golub}}, \bibinfo {author} {\bibfnamefont {I.}~\bibnamefont {Kuzmenko}},\
		and\ \bibinfo {author} {\bibfnamefont {Y.}~\bibnamefont {Avishai}},\
	}\bibfield  {title} {\bibinfo {title} {{Kondo Correlations and Majorana Bound
				States in a Metal to Quantum-Dot to Topological-Superconductor Junction}},\
	}\href {https://doi.org/10.1103/PhysRevLett.107.176802} {\bibfield  {journal}
		{\bibinfo  {journal} {Phys. Rev. Lett.}\ }\textbf {\bibinfo {volume} {107}},\
		\bibinfo {pages} {176802} (\bibinfo {year} {2011})}\BibitemShut {NoStop}%
	\bibitem [{\citenamefont {Lee}\ \emph {et~al.}(2013)\citenamefont {Lee},
		\citenamefont {Lim},\ and\ \citenamefont
		{L{\ifmmode\acute{o}\else\'{o}\fi}pez}}]{Lee2013Jun}%
	\BibitemOpen
	\bibfield  {author} {\bibinfo {author} {\bibfnamefont {M.}~\bibnamefont
			{Lee}}, \bibinfo {author} {\bibfnamefont {J.~S.}\ \bibnamefont {Lim}},\ and\
		\bibinfo {author} {\bibfnamefont {R.}~\bibnamefont
			{L{\ifmmode\acute{o}\else\'{o}\fi}pez}},\ }\bibfield  {title} {\bibinfo
		{title} {{Kondo effect in a quantum dot side-coupled to a topological
				superconductor}},\ }\href {https://doi.org/10.1103/PhysRevB.87.241402}
	{\bibfield  {journal} {\bibinfo  {journal} {Phys. Rev. B}\ }\textbf {\bibinfo
			{volume} {87}},\ \bibinfo {pages} {241402} (\bibinfo {year}
		{2013})}\BibitemShut {NoStop}%
	\bibitem [{\citenamefont {Cheng}\ \emph {et~al.}(2014)\citenamefont {Cheng},
		\citenamefont {Becker}, \citenamefont {Bauer},\ and\ \citenamefont
		{Lutchyn}}]{Cheng2014Sep}%
	\BibitemOpen
	\bibfield  {author} {\bibinfo {author} {\bibfnamefont {M.}~\bibnamefont
			{Cheng}}, \bibinfo {author} {\bibfnamefont {M.}~\bibnamefont {Becker}},
		\bibinfo {author} {\bibfnamefont {B.}~\bibnamefont {Bauer}},\ and\ \bibinfo
		{author} {\bibfnamefont {R.~M.}\ \bibnamefont {Lutchyn}},\ }\bibfield
	{title} {\bibinfo {title} {{Interplay between Kondo and Majorana Interactions
				in Quantum Dots}},\ }\href {https://doi.org/10.1103/PhysRevX.4.031051}
	{\bibfield  {journal} {\bibinfo  {journal} {Phys. Rev. X}\ }\textbf {\bibinfo
			{volume} {4}},\ \bibinfo {pages} {031051} (\bibinfo {year}
		{2014})}\BibitemShut {NoStop}%
	\bibitem [{\citenamefont {Aguado}(2017)}]{Aguado2017Oct}%
	\BibitemOpen
	\bibfield  {author} {\bibinfo {author} {\bibfnamefont {R.}~\bibnamefont
			{Aguado}},\ }\bibfield  {title} {\bibinfo {title} {{Majorana quasiparticles
				in condensed matter}},\ }\href {https://doi.org/10.1393/ncr/i2017-10141-9}
	{\bibfield  {journal} {\bibinfo  {journal} {La Rivista del Nuovo Cimento}\
		}\textbf {\bibinfo {volume} {40}},\ \bibinfo {pages} {523} (\bibinfo {year}
		{2017})}\BibitemShut {NoStop}%
	\bibitem [{\citenamefont {Kondo}(1964)}]{Kondo1964}%
	\BibitemOpen
	\bibfield  {author} {\bibinfo {author} {\bibfnamefont {J.}~\bibnamefont
			{Kondo}},\ }\bibfield  {title} {\bibinfo {title} {Resistance minimum in
			dilute magnetic alloys},\ }\href {https://doi.org/10.1143/PTP.32.37}
	{\bibfield  {journal} {\bibinfo  {journal} {Progress of Theoretical Physics}\
		}\textbf {\bibinfo {volume} {32}},\ \bibinfo {pages} {37} (\bibinfo {year}
		{1964})}\BibitemShut {NoStop}%
	\bibitem [{\citenamefont {Hewson}(1997)}]{Hewson1997}%
	\BibitemOpen
	\bibfield  {author} {\bibinfo {author} {\bibfnamefont {A.~C.}\ \bibnamefont
			{Hewson}},\ }\href {https://doi.org/10.1017/CBO9780511470752} {\emph
		{\bibinfo {title} {The Kondo problem to heavy fermions}}}\ (\bibinfo
	{publisher} {Cambridge University Press},\ \bibinfo {year}
	{1997})\BibitemShut {NoStop}%
	\bibitem [{\citenamefont {Goldhaber-Gordon}\ \emph {et~al.}(1998)\citenamefont
		{Goldhaber-Gordon}, \citenamefont {Shtrikman}, \citenamefont {Mahalu},
		\citenamefont {Abusch-Magder}, \citenamefont {Meirav},\ and\ \citenamefont
		{Kastner}}]{Goldhaber1998}%
	\BibitemOpen
	\bibfield  {author} {\bibinfo {author} {\bibfnamefont {D.}~\bibnamefont
			{Goldhaber-Gordon}}, \bibinfo {author} {\bibfnamefont {H.}~\bibnamefont
			{Shtrikman}}, \bibinfo {author} {\bibfnamefont {D.}~\bibnamefont {Mahalu}},
		\bibinfo {author} {\bibfnamefont {D.}~\bibnamefont {Abusch-Magder}}, \bibinfo
		{author} {\bibfnamefont {U.}~\bibnamefont {Meirav}},\ and\ \bibinfo {author}
		{\bibfnamefont {M.~A.}\ \bibnamefont {Kastner}},\ }\bibfield  {title}
	{\bibinfo {title} {Kondo effect in a single-electron transistor},\ }\href
	{https://doi.org/10.1038/34373} {\bibfield  {journal} {\bibinfo  {journal}
			{Nature}\ }\textbf {\bibinfo {volume} {391}},\ \bibinfo {pages} {156 EP }
		(\bibinfo {year} {1998})}\BibitemShut {NoStop}%
	\bibitem [{\citenamefont {Pustilnik}\ and\ \citenamefont
		{Glazman}(2001)}]{Pustilnik2001Nov}%
	\BibitemOpen
	\bibfield  {author} {\bibinfo {author} {\bibfnamefont {M.}~\bibnamefont
			{Pustilnik}}\ and\ \bibinfo {author} {\bibfnamefont {L.~I.}\ \bibnamefont
			{Glazman}},\ }\bibfield  {title} {\bibinfo {title} {{Kondo Effect in Real
				Quantum Dots}},\ }\href {https://doi.org/10.1103/PhysRevLett.87.216601}
	{\bibfield  {journal} {\bibinfo  {journal} {Phys. Rev. Lett.}\ }\textbf
		{\bibinfo {volume} {87}},\ \bibinfo {pages} {216601} (\bibinfo {year}
		{2001})}\BibitemShut {NoStop}%
	\bibitem [{\citenamefont {Vojta}\ \emph {et~al.}(2002)\citenamefont {Vojta},
		\citenamefont {Bulla},\ and\ \citenamefont {Hofstetter}}]{Vojta2002Apr}%
	\BibitemOpen
	\bibfield  {author} {\bibinfo {author} {\bibfnamefont {M.}~\bibnamefont
			{Vojta}}, \bibinfo {author} {\bibfnamefont {R.}~\bibnamefont {Bulla}},\ and\
		\bibinfo {author} {\bibfnamefont {W.}~\bibnamefont {Hofstetter}},\ }\bibfield
	{title} {\bibinfo {title} {{Quantum phase transitions in models of coupled
				magnetic impurities}},\ }\href {https://doi.org/10.1103/PhysRevB.65.140405}
	{\bibfield  {journal} {\bibinfo  {journal} {Phys. Rev. B}\ }\textbf {\bibinfo
			{volume} {65}},\ \bibinfo {pages} {140405} (\bibinfo {year}
		{2002})}\BibitemShut {NoStop}%
	\bibitem [{\citenamefont {Cornaglia}\ and\ \citenamefont
		{Grempel}(2005)}]{Cornaglia2005Feb}%
	\BibitemOpen
	\bibfield  {author} {\bibinfo {author} {\bibfnamefont {P.~S.}\ \bibnamefont
			{Cornaglia}}\ and\ \bibinfo {author} {\bibfnamefont {D.~R.}\ \bibnamefont
			{Grempel}},\ }\bibfield  {title} {\bibinfo {title} {{Strongly correlated
				regimes in a double quantum dot device}},\ }\href
	{https://doi.org/10.1103/PhysRevB.71.075305} {\bibfield  {journal} {\bibinfo
			{journal} {Phys. Rev. B}\ }\textbf {\bibinfo {volume} {71}},\ \bibinfo
		{pages} {075305} (\bibinfo {year} {2005})}\BibitemShut {NoStop}%
	\bibitem [{\citenamefont {{\ifmmode\check{Z}\else\v{Z}\fi}itko}\ and\
		\citenamefont {Bon{\ifmmode\check{c}\else\v{c}\fi}a}(2006)}]{Zitko2006Jan}%
	\BibitemOpen
	\bibfield  {author} {\bibinfo {author} {\bibfnamefont {R.}~\bibnamefont
			{{\ifmmode\check{Z}\else\v{Z}\fi}itko}}\ and\ \bibinfo {author}
		{\bibfnamefont {J.}~\bibnamefont {Bon{\ifmmode\check{c}\else\v{c}\fi}a}},\
	}\bibfield  {title} {\bibinfo {title} {{Enhanced conductance through
				side-coupled double quantum dots}},\ }\href
	{https://doi.org/10.1103/PhysRevB.73.035332} {\bibfield  {journal} {\bibinfo
			{journal} {Phys. Rev. B}\ }\textbf {\bibinfo {volume} {73}},\ \bibinfo
		{pages} {035332} (\bibinfo {year} {2006})}\BibitemShut {NoStop}%
	\bibitem [{\citenamefont {Chung}\ \emph {et~al.}(2008)\citenamefont {Chung},
		\citenamefont {Zarand},\ and\ \citenamefont
		{W{\ifmmode\ddot{o}\else\"{o}\fi}lfle}}]{Chung2008Jan}%
	\BibitemOpen
	\bibfield  {author} {\bibinfo {author} {\bibfnamefont {C.-H.}\ \bibnamefont
			{Chung}}, \bibinfo {author} {\bibfnamefont {G.}~\bibnamefont {Zarand}},\ and\
		\bibinfo {author} {\bibfnamefont {P.}~\bibnamefont
			{W{\ifmmode\ddot{o}\else\"{o}\fi}lfle}},\ }\bibfield  {title} {\bibinfo
		{title} {{Two-stage Kondo effect in side-coupled quantum dots: Renormalized
				perturbative scaling theory and numerical renormalization group analysis}},\
	}\href {https://doi.org/10.1103/PhysRevB.77.035120} {\bibfield  {journal}
		{\bibinfo  {journal} {Phys. Rev. B}\ }\textbf {\bibinfo {volume} {77}},\
		\bibinfo {pages} {035120} (\bibinfo {year} {2008})}\BibitemShut {NoStop}%
	\bibitem [{\citenamefont {Sasaki}\ \emph {et~al.}(2009)\citenamefont {Sasaki},
		\citenamefont {Tamura}, \citenamefont {Akazaki},\ and\ \citenamefont
		{Fujisawa}}]{Sasaki2009Dec}%
	\BibitemOpen
	\bibfield  {author} {\bibinfo {author} {\bibfnamefont {S.}~\bibnamefont
			{Sasaki}}, \bibinfo {author} {\bibfnamefont {H.}~\bibnamefont {Tamura}},
		\bibinfo {author} {\bibfnamefont {T.}~\bibnamefont {Akazaki}},\ and\ \bibinfo
		{author} {\bibfnamefont {T.}~\bibnamefont {Fujisawa}},\ }\bibfield  {title}
	{\bibinfo {title} {{Fano-Kondo Interplay in a Side-Coupled Double Quantum
				Dot}},\ }\href {https://doi.org/10.1103/PhysRevLett.103.266806} {\bibfield
		{journal} {\bibinfo  {journal} {Phys. Rev. Lett.}\ }\textbf {\bibinfo
			{volume} {103}},\ \bibinfo {pages} {266806} (\bibinfo {year}
		{2009})}\BibitemShut {NoStop}%
	\bibitem [{\citenamefont {Dias~da Silva}\ \emph {et~al.}(2013)\citenamefont
		{Dias~da Silva}, \citenamefont {Vernek}, \citenamefont {Ingersent},
		\citenamefont {Sandler},\ and\ \citenamefont {Ulloa}}]{DiasdaSilva2013May}%
	\BibitemOpen
	\bibfield  {author} {\bibinfo {author} {\bibfnamefont {L.~G. G.~V.}\
			\bibnamefont {Dias~da Silva}}, \bibinfo {author} {\bibfnamefont
			{E.}~\bibnamefont {Vernek}}, \bibinfo {author} {\bibfnamefont
			{K.}~\bibnamefont {Ingersent}}, \bibinfo {author} {\bibfnamefont
			{N.}~\bibnamefont {Sandler}},\ and\ \bibinfo {author} {\bibfnamefont {S.~E.}\
			\bibnamefont {Ulloa}},\ }\bibfield  {title} {\bibinfo {title}
		{{Spin-polarized conductance in double quantum dots: Interplay of Kondo,
				Zeeman, and interference effects}},\ }\href
	{https://doi.org/10.1103/PhysRevB.87.205313} {\bibfield  {journal} {\bibinfo
			{journal} {Phys. Rev. B}\ }\textbf {\bibinfo {volume} {87}},\ \bibinfo
		{pages} {205313} (\bibinfo {year} {2013})}\BibitemShut {NoStop}%
	\bibitem [{\citenamefont {W\'ojcik}\ and\ \citenamefont
		{Weymann}(2014)}]{Wojcik2014}%
	\BibitemOpen
	\bibfield  {author} {\bibinfo {author} {\bibfnamefont {K.~P.}\ \bibnamefont
			{W\'ojcik}}\ and\ \bibinfo {author} {\bibfnamefont {I.}~\bibnamefont
			{Weymann}},\ }\bibfield  {title} {\bibinfo {title} {Perfect spin polarization
			in t-shaped double quantum dots due to the spin-dependent fano effect},\
	}\href {https://doi.org/10.1103/PhysRevB.90.115308} {\bibfield  {journal}
		{\bibinfo  {journal} {Phys. Rev. B}\ }\textbf {\bibinfo {volume} {90}},\
		\bibinfo {pages} {115308} (\bibinfo {year} {2014})}\BibitemShut {NoStop}%
	\bibitem [{\citenamefont {W{\ifmmode\acute{o}\else\'{o}\fi}jcik}\ and\
		\citenamefont {Weymann}(2015)}]{Wojcik2015Apr}%
	\BibitemOpen
	\bibfield  {author} {\bibinfo {author} {\bibfnamefont {K.~P.}\ \bibnamefont
			{W{\ifmmode\acute{o}\else\'{o}\fi}jcik}}\ and\ \bibinfo {author}
		{\bibfnamefont {I.}~\bibnamefont {Weymann}},\ }\bibfield  {title} {\bibinfo
		{title} {{Two-stage Kondo effect in T-shaped double quantum dots with
				ferromagnetic leads}},\ }\href {https://doi.org/10.1103/PhysRevB.91.134422}
	{\bibfield  {journal} {\bibinfo  {journal} {Phys. Rev. B}\ }\textbf {\bibinfo
			{volume} {91}},\ \bibinfo {pages} {134422} (\bibinfo {year}
		{2015})}\BibitemShut {NoStop}%
	\bibitem [{\citenamefont {Wilson}(1975)}]{Wilson1975}%
	\BibitemOpen
	\bibfield  {author} {\bibinfo {author} {\bibfnamefont {K.~G.}\ \bibnamefont
			{Wilson}},\ }\bibfield  {title} {\bibinfo {title} {The renormalization group:
			Critical phenomena and the kondo problem},\ }\href
	{https://doi.org/10.1103/RevModPhys.47.773} {\bibfield  {journal} {\bibinfo
			{journal} {Rev. Mod. Phys.}\ }\textbf {\bibinfo {volume} {47}},\ \bibinfo
		{pages} {773} (\bibinfo {year} {1975})}\BibitemShut {NoStop}%
	\bibitem [{\citenamefont {Bulla}\ \emph {et~al.}(2008)\citenamefont {Bulla},
		\citenamefont {Costi},\ and\ \citenamefont {Pruschke}}]{Bulla2008}%
	\BibitemOpen
	\bibfield  {author} {\bibinfo {author} {\bibfnamefont {R.}~\bibnamefont
			{Bulla}}, \bibinfo {author} {\bibfnamefont {T.~A.}\ \bibnamefont {Costi}},\
		and\ \bibinfo {author} {\bibfnamefont {T.}~\bibnamefont {Pruschke}},\
	}\bibfield  {title} {\bibinfo {title} {Numerical renormalization group method
			for quantum impurity systems},\ }\href
	{https://doi.org/10.1103/RevModPhys.80.395} {\bibfield  {journal} {\bibinfo
			{journal} {Rev. Mod. Phys.}\ }\textbf {\bibinfo {volume} {80}},\ \bibinfo
		{pages} {395} (\bibinfo {year} {2008})}\BibitemShut {NoStop}%
	\bibitem [{\citenamefont {Weymann}\ and\ \citenamefont
		{W\'ojcik}(2017)}]{Wojcik2017}%
	\BibitemOpen
	\bibfield  {author} {\bibinfo {author} {\bibfnamefont {I.}~\bibnamefont
			{Weymann}}\ and\ \bibinfo {author} {\bibfnamefont {K.~P.}\ \bibnamefont
			{W\'ojcik}},\ }\bibfield  {title} {\bibinfo {title} {Transport properties of
			a hybrid majorana wire-quantum dot system with ferromagnetic contacts},\
	}\href {https://doi.org/10.1103/PhysRevB.95.155427} {\bibfield  {journal}
		{\bibinfo  {journal} {Phys. Rev. B}\ }\textbf {\bibinfo {volume} {95}},\
		\bibinfo {pages} {155427} (\bibinfo {year} {2017})}\BibitemShut {NoStop}%
	\bibitem [{\citenamefont {Flensberg}(2010)}]{Flensberg2010Nov}%
	\BibitemOpen
	\bibfield  {author} {\bibinfo {author} {\bibfnamefont {K.}~\bibnamefont
			{Flensberg}},\ }\bibfield  {title} {\bibinfo {title} {{Tunneling
				characteristics of a chain of Majorana bound states}},\ }\href
	{https://doi.org/10.1103/PhysRevB.82.180516} {\bibfield  {journal} {\bibinfo
			{journal} {Phys. Rev. B}\ }\textbf {\bibinfo {volume} {82}},\ \bibinfo
		{pages} {180516} (\bibinfo {year} {2010})}\BibitemShut {NoStop}%
	\bibitem [{\citenamefont {Hoffman}\ \emph {et~al.}(2017)\citenamefont
		{Hoffman}, \citenamefont {Chevallier}, \citenamefont {Loss},\ and\
		\citenamefont {Klinovaja}}]{Hoffman2017Jul}%
	\BibitemOpen
	\bibfield  {author} {\bibinfo {author} {\bibfnamefont {S.}~\bibnamefont
			{Hoffman}}, \bibinfo {author} {\bibfnamefont {D.}~\bibnamefont {Chevallier}},
		\bibinfo {author} {\bibfnamefont {D.}~\bibnamefont {Loss}},\ and\ \bibinfo
		{author} {\bibfnamefont {J.}~\bibnamefont {Klinovaja}},\ }\bibfield  {title}
	{\bibinfo {title} {{Spin-dependent coupling between quantum dots and
				topological quantum wires}},\ }\href
	{https://doi.org/10.1103/PhysRevB.96.045440} {\bibfield  {journal} {\bibinfo
			{journal} {Phys. Rev. B}\ }\textbf {\bibinfo {volume} {96}},\ \bibinfo
		{pages} {045440} (\bibinfo {year} {2017})}\BibitemShut {NoStop}%
	\bibitem [{\citenamefont {Meir}\ and\ \citenamefont
		{Wingreen}(1992)}]{Meir1992Apr}%
	\BibitemOpen
	\bibfield  {author} {\bibinfo {author} {\bibfnamefont {Y.}~\bibnamefont
			{Meir}}\ and\ \bibinfo {author} {\bibfnamefont {N.~S.}\ \bibnamefont
			{Wingreen}},\ }\bibfield  {title} {\bibinfo {title} {{Landauer formula for
				the current through an interacting electron region}},\ }\href
	{https://doi.org/10.1103/PhysRevLett.68.2512} {\bibfield  {journal} {\bibinfo
			{journal} {Phys. Rev. Lett.}\ }\textbf {\bibinfo {volume} {68}},\ \bibinfo
		{pages} {2512} (\bibinfo {year} {1992})}\BibitemShut {NoStop}%
	\bibitem [{NRG()}]{NRG_code}%
	\BibitemOpen
	\href@noop {} {\bibinfo  {journal} {We use the open-access Budapest Flexible
			DM-NRG code, http://www.phy.bme.hu/\~{}dmnrg/; O. Legeza, C. P. Moca, A. I.
			T\'{o}th, I. Weymann, G. Zar\'{a}nd, arXiv:0809.3143 (2008) (unpublished)}\
	}\BibitemShut {NoStop}%
	\bibitem [{\citenamefont {{\ifmmode\check{Z}\else\v{Z}\fi}itko}\ and\
		\citenamefont {Pruschke}(2009)}]{Zitko2009Feb}%
	\BibitemOpen
	\bibfield  {journal} {  }\bibfield  {author} {\bibinfo {author} {\bibfnamefont
			{R.}~\bibnamefont {{\ifmmode\check{Z}\else\v{Z}\fi}itko}}\ and\ \bibinfo
		{author} {\bibfnamefont {T.}~\bibnamefont {Pruschke}},\ }\bibfield  {title}
	{\bibinfo {title} {{Energy resolution and discretization artifacts in the
				numerical renormalization group}},\ }\href
	{https://doi.org/10.1103/PhysRevB.79.085106} {\bibfield  {journal} {\bibinfo
			{journal} {Phys. Rev. B}\ }\textbf {\bibinfo {volume} {79}},\ \bibinfo
		{pages} {085106} (\bibinfo {year} {2009})}\BibitemShut {NoStop}%
	\bibitem [{\citenamefont {Campo}\ and\ \citenamefont
		{Oliveira}(2005)}]{Campo2005Sep}%
	\BibitemOpen
	\bibfield  {author} {\bibinfo {author} {\bibfnamefont {V.~L.}\ \bibnamefont
			{Campo}}\ and\ \bibinfo {author} {\bibfnamefont {L.~N.}\ \bibnamefont
			{Oliveira}},\ }\bibfield  {title} {\bibinfo {title} {{Alternative
				discretization in the numerical renormalization-group method}},\ }\href
	{https://doi.org/10.1103/PhysRevB.72.104432} {\bibfield  {journal} {\bibinfo
			{journal} {Phys. Rev. B}\ }\textbf {\bibinfo {volume} {72}},\ \bibinfo
		{pages} {104432} (\bibinfo {year} {2005})}\BibitemShut {NoStop}%
	\bibitem [{\citenamefont {Freyn}\ and\ \citenamefont
		{Florens}(2009)}]{Freyn2009Mar}%
	\BibitemOpen
	\bibfield  {author} {\bibinfo {author} {\bibfnamefont {A.}~\bibnamefont
			{Freyn}}\ and\ \bibinfo {author} {\bibfnamefont {S.}~\bibnamefont
			{Florens}},\ }\bibfield  {title} {\bibinfo {title} {{Optimal broadening of
				finite energy spectra in the numerical renormalization group: Application to
				dissipative dynamics in two-level systems}},\ }\href
	{https://doi.org/10.1103/PhysRevB.79.121102} {\bibfield  {journal} {\bibinfo
			{journal} {Phys. Rev. B}\ }\textbf {\bibinfo {volume} {79}},\ \bibinfo
		{pages} {121102} (\bibinfo {year} {2009})}\BibitemShut {NoStop}%
	\bibitem [{\citenamefont {Weymann}\ and\ \citenamefont
		{Barna{\ifmmode\acute{s}\else\'{s}\fi}}(2013)}]{Weymann2013Aug}%
	\BibitemOpen
	\bibfield  {author} {\bibinfo {author} {\bibfnamefont {I.}~\bibnamefont
			{Weymann}}\ and\ \bibinfo {author} {\bibfnamefont {J.}~\bibnamefont
			{Barna{\ifmmode\acute{s}\else\'{s}\fi}}},\ }\bibfield  {title} {\bibinfo
		{title} {{Spin thermoelectric effects in Kondo quantum dots coupled to
				ferromagnetic leads}},\ }\href {https://doi.org/10.1103/PhysRevB.88.085313}
	{\bibfield  {journal} {\bibinfo  {journal} {Phys. Rev. B}\ }\textbf {\bibinfo
			{volume} {88}},\ \bibinfo {pages} {085313} (\bibinfo {year}
		{2013})}\BibitemShut {NoStop}%
	\bibitem [{\citenamefont {W{\ifmmode\acute{o}\else\'{o}\fi}jcik}\ and\
		\citenamefont {Weymann}(2019)}]{KWIW-2SQD_2019}%
	\BibitemOpen
	\bibfield  {author} {\bibinfo {author} {\bibfnamefont {K.~P.}\ \bibnamefont
			{W{\ifmmode\acute{o}\else\'{o}\fi}jcik}}\ and\ \bibinfo {author}
		{\bibfnamefont {I.}~\bibnamefont {Weymann}},\ }\bibfield  {title} {\bibinfo
		{title} {{Nonlocal pairing as a source of spin exchange and Kondo
				screening}},\ }\href {https://doi.org/10.1103/PhysRevB.99.045120} {\bibfield
		{journal} {\bibinfo  {journal} {Phys. Rev. B}\ }\textbf {\bibinfo {volume}
			{99}},\ \bibinfo {pages} {045120} (\bibinfo {year} {2019})}\BibitemShut
	{NoStop}%
	\bibitem [{\citenamefont {Bulla}\ and\ \citenamefont
		{Hewson}(1997)}]{Bulla1997Jun}%
	\BibitemOpen
	\bibfield  {author} {\bibinfo {author} {\bibfnamefont {R.}~\bibnamefont
			{Bulla}}\ and\ \bibinfo {author} {\bibfnamefont {A.~C.}\ \bibnamefont
			{Hewson}},\ }\bibfield  {title} {\bibinfo {title} {{Numerical renormalization
				group study of the O(3)-symmetric Anderson model}},\ }\href
	{https://doi.org/10.1007/s002570050458} {\bibfield  {journal} {\bibinfo
			{journal} {Z. Phys. B: Condens. Matter}\ }\textbf {\bibinfo {volume} {104}},\
		\bibinfo {pages} {333} (\bibinfo {year} {1997})}\BibitemShut {NoStop}%
	\bibitem [{\citenamefont {Bradley}\ \emph {et~al.}(1999)\citenamefont
		{Bradley}, \citenamefont {Bulla}, \citenamefont {Hewson},\ and\ \citenamefont
		{Zhang}}]{Bradley1999Oct}%
	\BibitemOpen
	\bibfield  {author} {\bibinfo {author} {\bibfnamefont {S.~C.}\ \bibnamefont
			{Bradley}}, \bibinfo {author} {\bibfnamefont {R.}~\bibnamefont {Bulla}},
		\bibinfo {author} {\bibfnamefont {A.~C.}\ \bibnamefont {Hewson}},\ and\
		\bibinfo {author} {\bibfnamefont {G.-M.}\ \bibnamefont {Zhang}},\ }\bibfield
	{title} {\bibinfo {title} {{Spectral densities of response functions for the
				O(3) symmetric Anderson and two channel Kondo models}},\ }\href
	{https://doi.org/10.1007/s100510051181} {\bibfield  {journal} {\bibinfo
			{journal} {Eur. Phys. J. B}\ }\textbf {\bibinfo {volume} {11}},\ \bibinfo
		{pages} {535} (\bibinfo {year} {1999})}\BibitemShut {NoStop}%
	\bibitem [{\citenamefont {W{\ifmmode\acute{o}\else\'{o}\fi}jcik}\ and\
		\citenamefont {Weymann}(2018)}]{KWIW-2SQD_2018}%
	\BibitemOpen
	\bibfield  {author} {\bibinfo {author} {\bibfnamefont {K.~P.}\ \bibnamefont
			{W{\ifmmode\acute{o}\else\'{o}\fi}jcik}}\ and\ \bibinfo {author}
		{\bibfnamefont {I.}~\bibnamefont {Weymann}},\ }\bibfield  {title} {\bibinfo
		{title} {{Interplay of the Kondo effect with the induced pairing in
				electronic and caloric properties of T-shaped double quantum dots}},\ }\href
	{https://doi.org/10.1103/PhysRevB.97.235449} {\bibfield  {journal} {\bibinfo
			{journal} {Phys. Rev. B}\ }\textbf {\bibinfo {volume} {97}},\ \bibinfo
		{pages} {235449} (\bibinfo {year} {2018})}\BibitemShut {NoStop}%
	\bibitem [{\citenamefont {Haldane}(1978)}]{Haldane1978Feb}%
	\BibitemOpen
	\bibfield  {author} {\bibinfo {author} {\bibfnamefont {F.~D.~M.}\
			\bibnamefont {Haldane}},\ }\bibfield  {title} {\bibinfo {title} {{Scaling
				Theory of the Asymmetric Anderson Model}},\ }\href
	{https://doi.org/10.1103/PhysRevLett.40.416} {\bibfield  {journal} {\bibinfo
			{journal} {Phys. Rev. Lett.}\ }\textbf {\bibinfo {volume} {40}},\ \bibinfo
		{pages} {416} (\bibinfo {year} {1978})}\BibitemShut {NoStop}%
	\bibitem [{\citenamefont {Guo}\ \emph {et~al.}(2020)\citenamefont {Guo},
		\citenamefont {Zhu}, \citenamefont {Zhou}, \citenamefont {Yu}, \citenamefont
		{Lu},\ and\ \citenamefont {Liang}}]{Guo2020Mar}%
	\BibitemOpen
	\bibfield  {author} {\bibinfo {author} {\bibfnamefont {X.}~\bibnamefont
			{Guo}}, \bibinfo {author} {\bibfnamefont {Q.}~\bibnamefont {Zhu}}, \bibinfo
		{author} {\bibfnamefont {L.}~\bibnamefont {Zhou}}, \bibinfo {author}
		{\bibfnamefont {W.}~\bibnamefont {Yu}}, \bibinfo {author} {\bibfnamefont
			{W.}~\bibnamefont {Lu}},\ and\ \bibinfo {author} {\bibfnamefont
			{W.}~\bibnamefont {Liang}},\ }\bibfield  {title} {\bibinfo {title} {{Gate
				tuning and universality of Two-stage Kondo effect in single molecule
				transistors}},\ }\href {https://arxiv.org/abs/2003.05346} {\bibfield
		{journal} {\bibinfo  {journal} {arXiv}\ } (\bibinfo {year} {2020})},\ \Eprint
	{https://arxiv.org/abs/2003.05346} {2003.05346} \BibitemShut {NoStop}%
	\bibitem [{\citenamefont {L{\ifmmode\acute{o}\else\'{o}\fi}pez}\ \emph
		{et~al.}(2014)\citenamefont {L{\ifmmode\acute{o}\else\'{o}\fi}pez},
		\citenamefont {Lee}, \citenamefont {Serra},\ and\ \citenamefont
		{Lim}}]{Lopez2014May}%
	\BibitemOpen
	\bibfield  {author} {\bibinfo {author} {\bibfnamefont {R.}~\bibnamefont
			{L{\ifmmode\acute{o}\else\'{o}\fi}pez}}, \bibinfo {author} {\bibfnamefont
			{M.}~\bibnamefont {Lee}}, \bibinfo {author} {\bibfnamefont {L.}~\bibnamefont
			{Serra}},\ and\ \bibinfo {author} {\bibfnamefont {J.~S.}\ \bibnamefont
			{Lim}},\ }\bibfield  {title} {\bibinfo {title} {{Thermoelectrical detection
				of Majorana states}},\ }\href {https://doi.org/10.1103/PhysRevB.89.205418}
	{\bibfield  {journal} {\bibinfo  {journal} {Phys. Rev. B}\ }\textbf {\bibinfo
			{volume} {89}},\ \bibinfo {pages} {205418} (\bibinfo {year}
		{2014})}\BibitemShut {NoStop}%
	\bibitem [{\citenamefont {Weymann}(2017)}]{Weymann2017Jan}%
	\BibitemOpen
	\bibfield  {author} {\bibinfo {author} {\bibfnamefont {I.}~\bibnamefont
			{Weymann}},\ }\bibfield  {title} {\bibinfo {title} {{Spin Seebeck effect in
				quantum dot side-coupled to topological superconductor}},\ }\href
	{https://doi.org/10.1088/1361-648x/aa5526} {\bibfield  {journal} {\bibinfo
			{journal} {J. Phys.: Condens. Matter}\ }\textbf {\bibinfo {volume} {29}},\
		\bibinfo {pages} {095301} (\bibinfo {year} {2017})}\BibitemShut {NoStop}%
\end{thebibliography}

%

\end{document}